%% file: main.tex
\PassOptionsToPackage{dvipsnames}{xcolor}
\documentclass[manuscript,nonacm]{acmart}

\input{package}

\input{macro}
\showrevisionsfalse

\title{Bridging Generation and Training: A Systematic Review of Quality Issues in LLMs for Code}

\author{Kaifeng He}
\authornote{These authors contributed equally to this work.}
\email{hekf5@mail2.sysu.edu.cn}
\affiliation{%
  \institution{Sun Yat-sen University}
  \city{Zhuhai}
  \country{China}
}

\author{Xiaojun Zhang}
\authornotemark[1]
\email{zhangxj229@mail2.sysu.edu.cn}
\affiliation{%
  \institution{Sun Yat-sen University}
  \city{Zhuhai}
  \country{China}
}

\author{Peiliang Cai}
\authornotemark[1]
\email{caipliang3@mail2.sysu.edu.cn}
\affiliation{%
  \institution{Sun Yat-sen University}
  \city{Zhuhai}
  \country{China}
}

\author{Mingwei Liu}
\authornote{Corresponding author.}
\email{liumw26@mail.sysu.edu.cn}
\affiliation{%
  \institution{Sun Yat-sen University}
  \city{Zhuhai}
  \country{China}
}

\author{Yanlin Wang}
\email{wangylin36@mail.sysu.edu.cn}
\affiliation{%
  \institution{Sun Yat-sen University}
  \city{Zhuhai}
  \country{China}
}

\author{Chong Wang}
\email{chong.wang@ntu.edu.sg}
\affiliation{%
  \institution{Nanyang Technological University}
  \city{Singapore}
  \country{Singapore}
}

\author{Kaifeng Huang}
\email{kaifengh@tongji.edu.cn}
\affiliation{%
  \institution{Tongji University}
  \city{Shanghai}
  \country{China}
}

\author{Bihuan Chen}
\email{bhchen@fudan.edu.cn}
\affiliation{%
  \institution{Fudan University}
  \city{Shanghai}
  \country{China}
}

\author{Xin Peng}
\email{pengxin@fudan.edu.cn}
\affiliation{%
  \institution{Fudan University}
  \city{Shanghai}
  \country{China}
}

\author{Zibin Zheng}
\email{zhzibin@mail.sysu.edu.cn}
\affiliation{%
  \institution{Sun Yat-sen University}
  \city{Zhuhai}
  \country{China}
}

\renewcommand{\shortauthors}{K. He, X. Zhang, P. Cai, M. Liu, Y. Wang, C. Wang, K. Huang, B. Chen, X. Peng, and Z. Zheng}

\begin{document}

\input{sections/abstract}



\maketitle

\input{sections/introduction}
\input{sections/background}
\input{sections/methodology}
\input{sections/findings}
\input{sections/discussion}
\input{sections/threats}
\input{sections/conclusion}

\bibliographystyle{ACM-Reference-Format}
\balance
\bibliography{ref}

\end{document}

%% file: package.tex
\usepackage[utf8]{inputenc}
\usepackage[T1]{fontenc}
\usepackage[english]{babel}
\usepackage{xspace}
\usepackage{pifont}

\usepackage{amsmath,amssymb}
\usepackage{array}

\usepackage{graphicx}

\usepackage[framemethod=TikZ]{mdframed}

\usepackage{balance}

\usepackage{silence}
\usepackage[edges]{forest}
\useforestlibrary{edges}
\usepackage{varwidth}

%% file: macro.tex
\newcommand{\llm}{LLM\xspace}
\newcommand{\llms}{LLMs\xspace}

\newcommand{\numstudies}{$114$\xspace}

\newif\ifshowrevisions
\showrevisionstrue
\definecolor{RevisionBlue}{RGB}{0,76,153}

\DeclareRobustCommand{\rev}[1]{%
    \ifshowrevisions
        \textcolor{blue}{#1}%
    \else
        #1%
    \fi
}

\newenvironment{revisionblock}{%
    \ifshowrevisions\color{blue}\fi
}{}

\newcommand{\CodeIssue}{Generated Code Quality Issue\xspace}

\newcommand{\DataIssue}{Training Data Quality Issue\xspace}
\newcommand{\LLmodels}{Large language models\xspace}

\newcommand{\codeissues}{generated code quality issues\xspace}

\newcommand{\dataissues}{training data quality issues\xspace}

\newenvironment{rqsummary}[1][Summary and Insights]{
    \begin{mdframed}[
        linewidth=2pt,
        linecolor=black!60,
        topline=false,
        rightline=false,
        bottomline=false,
        backgroundcolor=gray!10,
        roundcorner=2pt,
        innertopmargin=8pt,
        innerbottommargin=8pt,
        innerleftmargin=10pt,
        innerrightmargin=10pt,
        skipabove=1em,
        skipbelow=1em
    ]
    \textbf{\textit{#1.}}\quad
}{
    \end{mdframed}
}

%% file: sections/abstract.tex
\begin{abstract}
\begin{revisionblock}
\LLmodels (\llms) frequently generate defective code, ranging from logical bugs to security vulnerabilities. Training-data quality is one contributor to these outcomes alongside model architecture, optimization and alignment, prompting and inference context, and decoding. However, the literature connecting specific \dataissues to observable \codeissues has not been synthesized systematically. We review \numstudies included papers and establish taxonomies of eight generated-code quality dimensions and two broad groups of training-data issues. We then synthesize 12 recurring relationship families between training-data issues and generated-code outcomes, tracing each to its supporting papers, evidence basis, and limitations. We also synthesize detection and mitigation techniques across the data, model, and generation lifecycle. The findings position data curation as one layer of code-generation quality assurance coordinated with model- and inference-level controls, and identify provenance and controlled attribution as open research needs. Our repository is available at \url{https://github.com/SYSUSELab/From-Data-to-Code}.
\end{revisionblock}
\end{abstract}

%% file: sections/introduction.tex
\section{Introduction}
\label{sec:introduction}
\begin{revisionblock}
Large language models (\llms) have demonstrated strong performance across a range of application domains~\cite{chang2024survey}. Transformer-based models, including models specialized for code, are increasingly used in software development~\cite{102}. They have been applied to code completion~\cite{alkaswan2024completion}, program repair~\cite{fan2023automated,li2026combenchrepolevelrealworldbenchmark}, test-case generation~\cite{yuan2024evaluating}, and code documentation~\cite{geng2024large}. A study of GitHub Copilot in collaborative open-source development reports productivity gains following its adoption~\cite{song2024impact}.
\end{revisionblock}

\input{figures/LLMGeneratedCode}
\begin{revisionblock}
Despite these benefits, \llm-generated code can contain quality issues that limit reliable use~\cite{10}. Studies report syntax and logic errors, security weaknesses, and maintainability concerns in generated code~\cite{chen2024survey}. Figure~\ref{fig:LLMGeneratedCode} illustrates a Python/Flask registration endpoint with three overlapping defects: an unparameterized SQL query (Line 11) that permits SQL injection, use of the deprecated \texttt{pandas.append()} method (Line 13) that breaks compatibility with modern pandas versions, and database operations without exception handling (Lines 16--19) that may result in runtime failure. Such defects increase the need for systematic verification before generated code is integrated into software systems, particularly in security-sensitive settings~\cite{negri2024systematic}.
\end{revisionblock}

\input{figures/TrainingDataCode}
\begin{revisionblock}
Training data is one component of an \llm code-generation system. Studies report relationships between corpus defects or coverage and generated-code outcomes, and controlled data interventions show that filtering, rewriting, or changing data composition can alter quality measures in particular settings~\cite{3,63,76,115}. Corpus audits, cross-model comparisons, and authors' analyses provide additional evidence when training sets are undisclosed. Figure~\ref{fig:TrainingDataCode} illustrates the type of association examined: both snippets use the obsolete Python~2 \texttt{urllib2} API in different tasks. Such associations must be interpreted alongside architecture, optimization and alignment, prompting and retrieved context, decoding, and deployment tooling.
\end{revisionblock}

\begin{revisionblock}
Research on detection and mitigation is distributed across data curation, model training, and post-generation refinement~\cite{huynh2025large,pysmennyi2025ai}. Existing studies commonly evaluate individual techniques or lifecycle stages, leaving limited synthesis of how controls can be combined across settings. Against this background, we focus on three gaps:
\end{revisionblock}
\ding{182} \rev{Fragmented quality perspectives: Existing work includes broad surveys of \llm code generation, studies of individual quality concerns such as security vulnerabilities, and data-focused analyses~\cite{114,139,wang2023data}. However, these strands do not consistently synthesize \codeissues and \dataissues through a shared software engineering quality perspective;}
\ding{183} \rev{Limited evidence on relationships between training-data and generated-code quality: While large-scale model reports acknowledge the opacity between data curation and output quality~\cite{StarCoder}, few studies isolate a specific dataset property while holding model and inference factors constant, and undisclosed provenance prevents reliable tracing from an output to its training sources;}
\ding{184} \rev{Limited cross-stage synthesis of detection and mitigation methods: Existing reviews catalogue techniques used at different lifecycle stages, but provide limited comparison of how these controls can be combined across data preparation, model development, and generated-code validation~\cite{hou2024large}.}

\begin{revisionblock}
To address these gaps, this paper presents a systematic literature review of \numstudies included papers on \llm training data and code generation. We organize \codeissues into eight dimensions and \dataissues into code attributes and non-code or corpus-level attributes. We identify 12 recurring relationship families, trace each to its supporting papers and evidence, and show where research attention is concentrated. Together with the reviewed detection and mitigation methods, this synthesis positions data curation within end-to-end quality assurance spanning model training, inference-time controls, and software-engineering validation.
\end{revisionblock}

In summary, the main contributions of this work are as follows:

\begin{itemize}

\item We \rev{synthesize} a \textbf{dual-dimensional taxonomy} that categorizes \codeissues across \rev{eight} quality dimensions \rev{grounded in established software-quality attributes~\cite{iso25010} and adapted to \llm-generated code}, and organizes \dataissues into two major groups.

\item \rev{We develop a \textbf{literature-grounded mapping framework} of 12 recurring relationships between \dataissues and observed \codeissues, trace each relationship to its supporting papers, and summarize its evidence and limitations.}

\item We synthesize \textbf{existing detection and mitigation techniques} across the lifecycle of \llms for code generation, covering data curation, model training, and post-generation validation.

\end{itemize}

The remainder of this paper is organized as follows:
Section~\ref{sec:background} provides background on \llms for code generation, generated code quality, and dataset quality.
Section~\ref{sec:methodology} details our systematic review methodology.
Section~\ref{sec:findings} presents our core findings addressing the three research gaps, with a focus on \rev{relationships between training-data and generated-code quality} and integrated detection/mitigation frameworks.
Section~\ref{sec:discussion} discusses key insights, challenges, and implications for data-centric quality assurance.
Section~\ref{sec:threats} analyzes threats to validity.
Section~\ref{sec:conclusion} concludes with a summary and future work.

%% file: figures/LLMGeneratedCode.tex
\begin{figure}[htbp]
    \centering
    \includegraphics[width=0.7\linewidth]{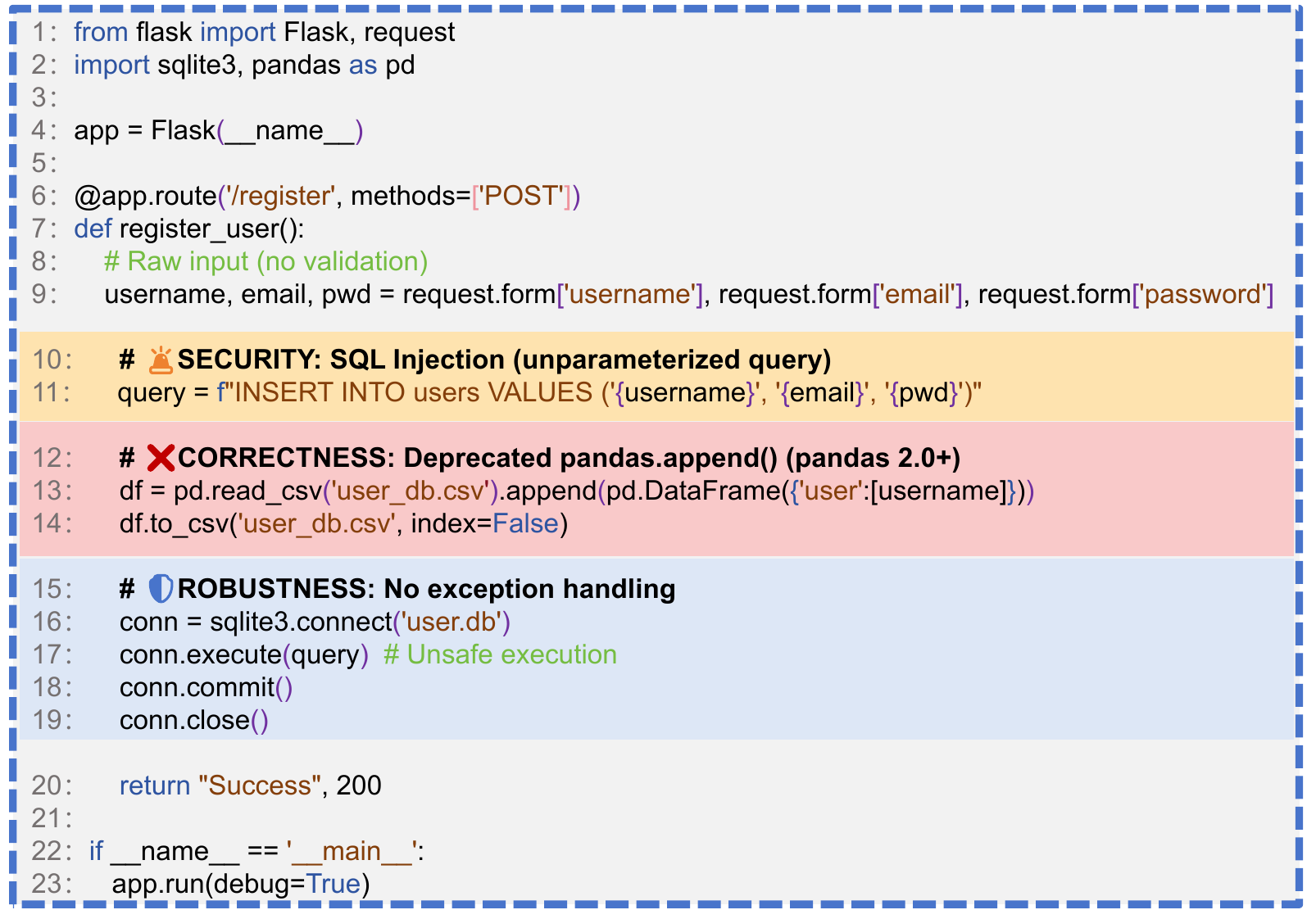}
    \caption{\rev{Illustrative Python/Flask registration endpoint showing three overlapping output defects. The highlighted regions identify an unparameterized SQL statement at Line~11 (security: SQL injection), a deprecated \texttt{DataFrame.append()} call at Line~13 (correctness and compatibility), and database operations without exception handling at Lines~16--19 (robustness). The artifact therefore receives multiple quality labels.}}
    \Description{LLM-generated code with multiple quality defects}
    \label{fig:LLMGeneratedCode}
\end{figure}

%% file: figures/TrainingDataCode.tex
\begin{figure}[htbp]
    \centering
    \includegraphics[width=0.7\linewidth]{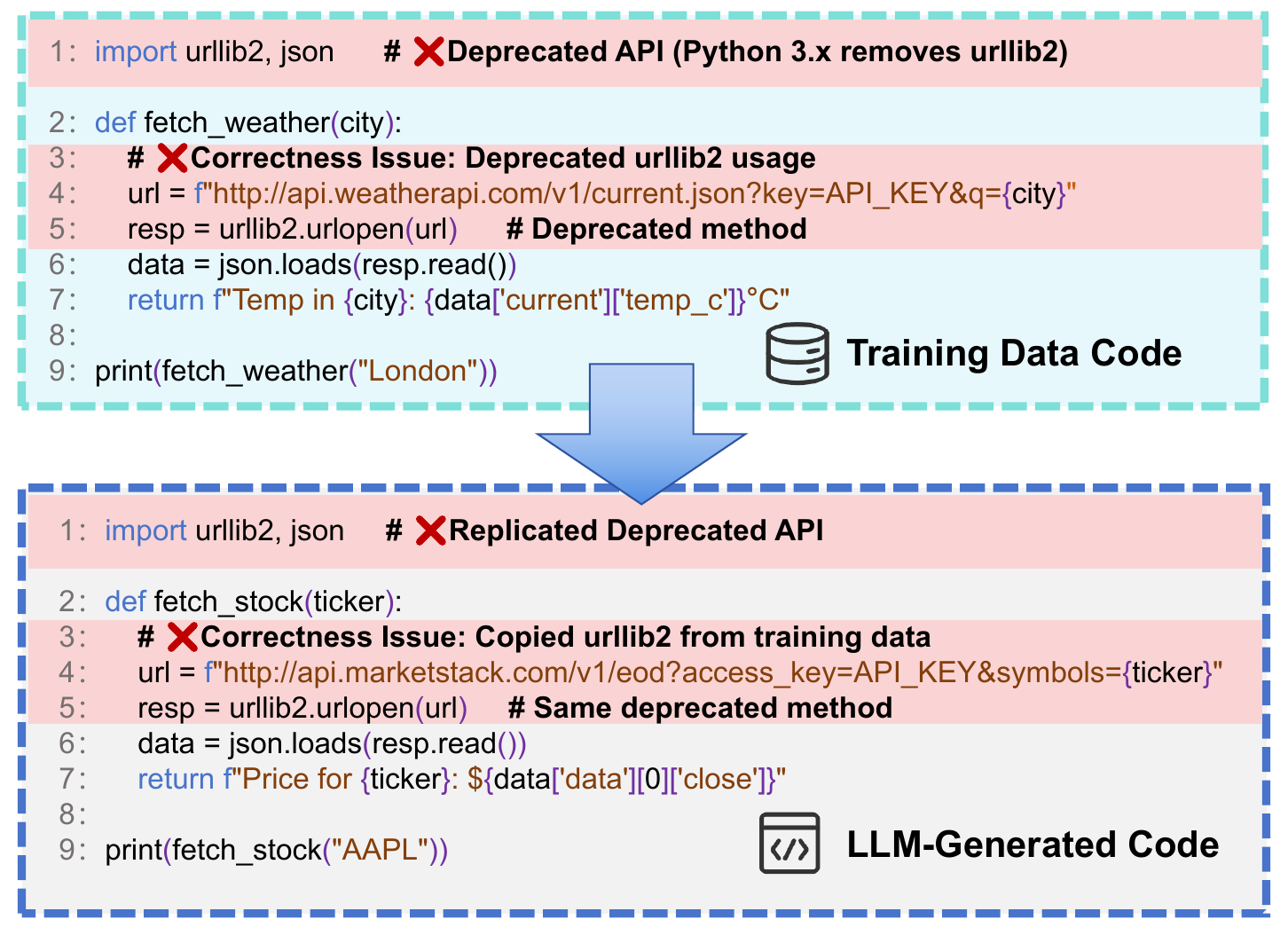}
    \caption{\rev{Illustrative example of an obsolete API pattern appearing in both training and generated code. The upper snippet uses the Python~2 \texttt{urllib2} API, which is unavailable in Python~3; the lower snippet performs a different task using the same obsolete API. The highlighted lines mark the shared pattern.}}
    \Description{An obsolete API pattern shown in a training example and in generated code for a different task.}
    \label{fig:TrainingDataCode}
\end{figure}

%% file: sections/background.tex
\section{Background}
\label{sec:background}

This section provides the conceptual background for the subsequent review. We first introduce \llms in code generation to clarify the technical context in which the reviewed papers are situated. We then describe the data-driven lifecycle of LLM-based code generation, because this lifecycle provides the basis for understanding \rev{reported relationships between \dataissues and generated code}. Next, we define \codeissues and \dataissues from a dual perspective to establish a consistent terminology for the later synthesis. Finally, we review representative surveys to position this study in relation to existing review efforts and to clarify the gap addressed in this work.

\subsection{\llms in Code Generation}
\begin{revisionblock}
Transformer-based \llms are used for generating, understanding, and reasoning about source code~\cite{zhang2024surveylargelanguagemodels,vaswani2023attentionneed}. They include general-purpose models such as GPT-4~\cite{80}, Gemini~\cite{team2023gemini}, and Qwen~\cite{4}, as well as code-specialized models such as Codex~\cite{102}, CodeT5~\cite{CodeT5}, DeepSeek-Coder~\cite{79}, Code Llama~\cite{101}, and StarCoder~\cite{StarCoder}. Code-specialized models learn programming-language patterns from large code corpora, while the composition and disclosure of training data vary across model families.
\end{revisionblock}

\begin{revisionblock}
Compared with natural-language generation, code generation must satisfy exact constraints on syntax, semantics, and executability~\cite{allamanis2018surveymachinelearningbig}. Small syntactic deviations can prevent compilation, while semantic inconsistencies can produce incorrect behavior. Although \llms support applications including automatic bug fixing, code translation, and test synthesis~\cite{zhang2024surveylargelanguagemodels}, they can produce outputs that are structurally plausible but functionally incorrect. Such outputs are sometimes described as ``hallucinated code''~\cite{zhang2025sirenssongaiocean}. This gap between surface-level fluency and software reliability motivates systematic evaluation of code generation quality.
\end{revisionblock}

\subsection{Data-Driven Lifecycle of LLM-based Code Generation}

\rev{Understanding factors associated with degraded generation quality~\cite{liu2023codegeneratedchatgptreally} requires examining the lifecycle of \llms in code generation tasks. As illustrated in Figure~\ref{fig:lifecycle}, this lifecycle proceeds from Raw Data through Data Processing, Model Training, and Code Generation.}

\input{figures/lifecycle}

\begin{revisionblock}
Training data shapes the statistical and semantic patterns available to \llms for code~\cite{lee2022deduplicatingtrainingdatamakes}. Model training typically includes two broad phases~\cite{ouyang2022traininglanguagemodelsfollow}. Large-scale pre-training draws on heterogeneous code corpora and documentation, which provide broad programming knowledge while also containing noise, deprecated APIs, and insecure coding practices~\cite{139}. Instruction tuning or reinforcement learning with human feedback then uses curated examples and preferences to align the model with human intent~\cite{gunasekar2023textbooksneed}. The resulting behavior reflects these data sources together with the architecture and optimization applied to them.
\end{revisionblock}

\begin{revisionblock}
Code generation remains connected to this data lineage because learned distributions influence which patterns a model reproduces or generalizes. Unmanaged corpus flaws can therefore accumulate as data-driven technical debt~\cite{hiddentechnicaldebt}. The ``Mappings'' module in Figure~\ref{fig:lifecycle} represents relationships reported between training-data issues, such as imbalance, vulnerabilities, or redundancy, and generated-code outcomes, such as incorrectness, unreadability, or insecurity. Architecture, optimization and alignment, prompts and retrieved context, decoding, and deployment controls act alongside the training data. The right side of the figure accordingly places mitigation at the data, model, and generation levels.
\end{revisionblock}

\subsection{Quality Issues: A Dual-Perspective Definition}
\label{sec:bgDefi}

\rev{Traditional software quality frameworks (e.g., ISO/IEC 25010~\cite{iso25010}) provide an established software engineering reference for evaluating attributes such as functional behavior, reliability, security, performance efficiency, and maintainability. \llm-generated code must meet these quality requirements, while task-conditioned generation also raises response-level concerns alongside established software attributes.} Given the fragmented terminology currently used to describe these emergent issues, we first formalize the two core concepts investigated in this review to establish a consistent vocabulary.

\begin{revisionblock}
First, we define a \textbf{\CodeIssue} as any anomaly in \llm-generated code that degrades its functional correctness, non-functional quality, risk-related security and compliance, or generation-specific attributes. The unit of analysis is the generated output artifact; intrinsic model metrics fall outside this scope. These issues range from traditional syntax and logic errors to generation-specific anomalies such as hallucinated APIs and unusable repetitive responses. Their origins may involve interactions among training data, model architecture and optimization, alignment, and inference conditions~\cite{12,28,he2019quantifying}.

Second, we define a \textbf{\DataIssue} as any defect or distributional bias within pre-training or fine-tuning corpora used for code generation. These issues include flaws in individual samples, such as incorrect, deprecated, or unmaintainable code and noisy text, together with corpus-level properties such as language imbalance, redundancy, inadequate diversity, benchmark contamination, and low-value content~\cite{lee2022deduplicatingtrainingdatamakes,matton2024leakagecodegenerationevaluation}. Such conditions can alter the programming patterns and associations learned by an \llm.

The relationships reported between \dataissues and \codeissues remain fragmented across software engineering and machine-learning research~\cite{gao2024currentchallengessoftwareengineering}. A shared vocabulary supports their systematic synthesis and informs quality controls across data preparation, model training, and code generation.
\end{revisionblock}

\subsection{Related Surveys}
\label{sec:related_surveys}

Recent literature has extensively explored \llms from either the data management perspective or the code generation quality perspective.

From the data perspective, existing work tends to isolate specific phases of the model lifecycle. For pre-training, researchers have systematized data processing pipelines, detailing mechanisms like data deduplication, quality filtering, and domain mixing~\cite{zhou2025surveyllmtimesdata}. For post-training, efforts focus on data-efficient paradigms---such as data selection, distillation, and synthetic data generation---to mitigate high annotation costs and diminishing returns from data scaling~\cite{Luo_2025}.

\rev{Conversely, work from the code generation quality perspective ranges from broad surveys of \llm code generation~\cite{114} to focused evaluations of particular output-quality attributes. Security-oriented surveys identify vulnerabilities in AI-generated code and review verification and mitigation approaches~\cite{negri2024systematic}.} Extending beyond security, empirical studies like Kharma et al.~\cite{65} conduct multi-dimensional evaluations across diverse programming languages, revealing pronounced architectural and memory-safety degradation in low-level languages like C/C++. Furthermore, recent work highlights non-functional discrepancies, demonstrating that while \llm-generated code achieves baseline readability, it exhibits significant stylistic inconsistencies in API preferences and formatting when compared to human developers~\cite{130}.

\rev{Table~\ref{tab:related_survey_comparison} summarizes the scope of treatment in representative reviews across these complementary perspectives.}
\input{tables/related_survey_comparison}

\rev{Taken together, the representative reviews in Table~\ref{tab:related_survey_comparison} cover complementary subsets of the \llm-for-code lifecycle, including training-data quality, generated-code quality, and quality-assurance methods. This review connects these areas by synthesizing reported relationships between data and code quality and by organizing detection and mitigation across the data, model, and generation stages.}

%% file: figures/lifecycle.tex
\begin{figure}[htbp]
    \centering
    \includegraphics[width=0.9\linewidth]{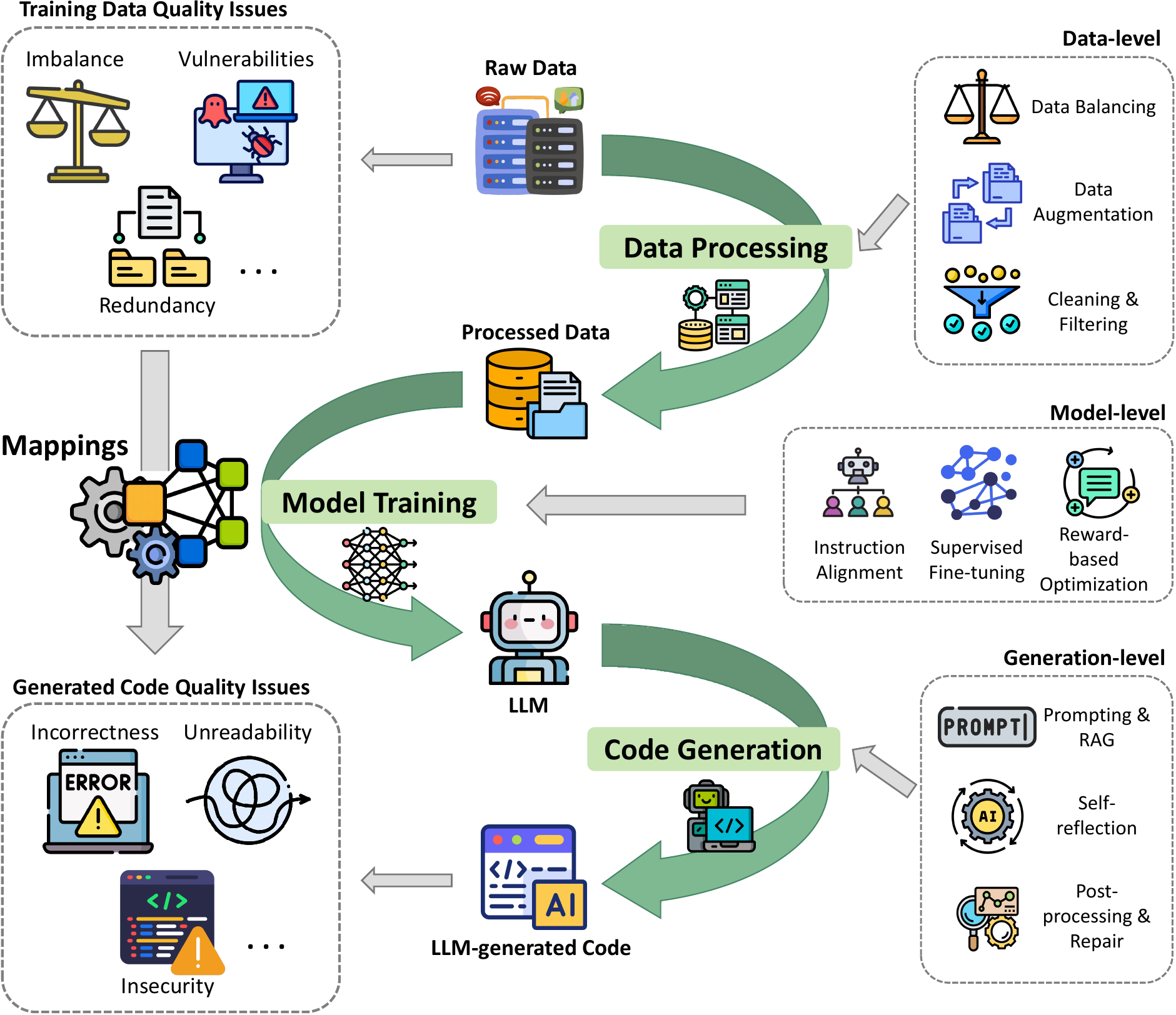}
    \caption{\rev{Author-developed conceptual overview of quality issues and intervention points across the lifecycle of an \llm for code. Green arrows trace raw data through data processing, processed data, model training, the \llm, code generation, and generated code. The boxes on the left show example training-data and generated-code issue classes connected by the RQ3 mappings; the boxes on the right locate data-, model-, and generation-level mitigation families. The mapping arrow represents relationships between these issue classes synthesized from the reviewed literature.}}
    \Description{Conceptual Framework of Quality Issues and Mitigation in the LLM Lifecycle}
    \label{fig:lifecycle}
\end{figure}

%% file: tables/related_survey_comparison.tex
\begin{table*}[t]
\centering
\caption{\rev{Scope of treatment in representative reviews of \llms for code and data. Columns identify whether a review explicitly synthesizes training-data issues, generated-code issues, reported relationships between them, and quality-assurance approaches. Entries describe the stated synthesis scope; ``Not synthesized'' means that the topic is not an explicit synthesis target.}}
\label{tab:related_survey_comparison}
\small
\renewcommand{\arraystretch}{1.15}
\setlength{\tabcolsep}{2.5pt}
\begin{tabular}{@{}>{\raggedright\arraybackslash}p{2.35cm} >{\raggedright\arraybackslash}p{2.85cm} >{\raggedright\arraybackslash}p{2.85cm} >{\raggedright\arraybackslash}p{2.85cm} >{\raggedright\arraybackslash}p{2.85cm}@{}}
\toprule
\multicolumn{1}{c}{\rev{\textbf{\shortstack{Representative\\review}}}} &
\multicolumn{1}{c}{\rev{\textbf{\shortstack{Training-data\\treatment}}}} &
\multicolumn{1}{c}{\rev{\textbf{\shortstack{Generated-code\\treatment}}}} &
\multicolumn{1}{c}{\rev{\textbf{\shortstack{Data--code\\relationship}}}} &
\multicolumn{1}{c}{\rev{\textbf{\shortstack{Quality-assurance\\treatment}}}} \\
\midrule
\rev{Hou et al.~\cite{hou2024large}}\newline
\rev{\textit{LLMs for SE}} &
\rev{SE datasets: collection, preprocessing, and use} &
\rev{SE tasks and \llm performance} &
\rev{Not synthesized} &
\rev{Model optimization and performance evaluation} \\

\rev{Jiang et al.~\cite{114}}\newline
\rev{\textit{Code generation}} &
\rev{Code-data curation} &
\rev{Code-generation developments and performance evaluation} &
\rev{Not synthesized} &
\rev{Benchmark-based performance evaluation} \\

\rev{Luo et al.~\cite{Luo_2025}}\newline
\rev{\textit{Data-centric post-training}} &
\rev{Data selection and quality enhancement} &
\rev{Not code-specific} &
\rev{Not synthesized} &
\rev{Post-training data selection and refinement} \\

\rev{Negri-Ribalta et al.~\cite{negri2024systematic}}\newline
\rev{\textit{Code security}} &
\rev{Dataset curation as security mitigation} &
\rev{Security flaws and exploits} &
\rev{Discussed in mitigation; not mapped} &
\rev{Security verification and mitigation} \\

\rev{Abdollahi et al.~\cite{abdollahi2026benchmark}}\newline
\rev{\textit{Code-intelligence benchmarks}} &
\rev{Not synthesized (evaluation data only)} &
\rev{Benchmark coverage of code-generation tasks} &
\rev{Not synthesized} &
\rev{Benchmark construction, contamination checks, and evaluation protocols} \\
\midrule
\rev{\textbf{This review}} &
\rev{\textbf{Training-data quality issues}} &
\rev{\textbf{Generated-code quality issues}} &
\rev{\textbf{Reported data--output relationships}} &
\rev{\textbf{Lifecycle-wide detection and mitigation}} \\
\bottomrule
\end{tabular}
\Description{Scope of treatment in representative reviews of LLMs for code and data, including training-data issues, generated-code issues, reported relationships between them, and quality-assurance approaches.}
\end{table*}

%% file: sections/methodology.tex
\section{Review Methodology}
\label{sec:methodology}

We follow established guidelines for conducting software engineering systematic literature reviews~\cite{kitchenham2007guidelines}. \rev{The paper-collection and filtering process, depicted in Figure~\ref{fig:paper_collection}, comprises seven stages: defining the research questions, designing the search strings, searching the databases, consolidating and de-duplicating the retrieved records, applying the inclusion and exclusion criteria, conducting backward and forward snowballing, and assessing study quality.}

\input{figures/paper_collection}

\subsection{Research Questions}
\label{sec:RQs}
This survey systematically investigates \codeissues in \llm-generated code and their potential origins in training datasets. To structure this research problem, we formulate the following five research questions (RQs).

\begin{itemize}
\item \textbf{RQ1: What \codeissues exist in \llm-generated code, and how can they be systematically categorized?}
This question identifies and organizes the spectrum of quality issues in generated code, establishing a taxonomy of \codeissues.

\item \textbf{RQ2: What \dataissues exist in datasets used for \llm training, and how can they be categorized?}
This question examines \dataissues in training datasets and organizes them into a structured taxonomy.

\item \rev{\textbf{RQ3: What relationships between \dataissues and \codeissues are reported in the literature?}}
\rev{This question synthesizes how the included papers relate \dataissues to \codeissues and summarizes the evidence and limitations for each relationship.}

\item \textbf{RQ4: What techniques exist for detecting \codeissues and \dataissues?}
This question reviews existing methods for identifying quality issues in both generated code and training data.

\item \textbf{RQ5: What mitigation strategies have been proposed to address these quality issues?}
This question synthesizes existing mitigation approaches into a lifecycle-oriented framework for improving the quality of \llm-based code generation systems.
\end{itemize}

Overall, these RQs are organized to progressively move from problem characterization to \rev{relationship synthesis} and finally to issue handling. RQ1 and RQ2 first characterize the quality issues in generated code and training datasets, respectively. \rev{Building on this foundation, RQ3 synthesizes these reported relationships into 12 families.} Finally, RQ4 and RQ5 focus on issue handling by reviewing existing detection techniques and mitigation strategies across the \llm lifecycle.

\subsection{Search Strategy}
\label{sec:search}

We employed a multi-stage search strategy combining systematic keyword-based querying with backward and forward snowballing~\cite{wohlin2014snowballing}. The common Boolean structure comprised three conceptual groups:
\begin{enumerate}
    \item \textbf{\llms for Code:} \rev{\texttt{((``large language model*'' OR ``LLM*'' OR ``code LLM*'') AND (``code generation'' OR ``code completion'' OR ``program repair''))}}
    \item \textbf{Generated Code Quality Issues:} \texttt{AND (``code quality'' OR ``defect'' OR ``bug'' OR ``vulnerability'' OR ``hallucination'' OR ``correctness'' OR ``security'' OR ``compliance'' OR ``robustness'' OR ``maintainability'' OR ``understandability'' OR ``efficiency'' OR ``repetition'')}
    \item \textbf{Training Data Quality Issues:} \texttt{AND (``training data'' OR ``dataset'' OR ``corpus'' OR ``data quality'' OR ``data noise'' OR ``data contamination'' OR ``data leakage'' OR ``imbalance'' OR ``redundancy'' OR ``duplication'' OR ``diversity'')}
\end{enumerate}

\rev{We adapted the field syntax to each database while retaining the Boolean structure above. We limited the database searches to records published through December~31, 2025; they returned 2,298 records in total.}

The initial database search was executed across five major digital libraries representing software engineering and artificial intelligence venues:

\begin{itemize}
    \item \textbf{IEEE Xplore:} Captures works published in venues such as \textit{IEEE Transactions on Software Engineering (TSE)}, \textit{ICSE}, and \textit{ASE}. \rev{(Search scope: \textit{All Metadata}; 380 records).}
    \item \textbf{ACM Digital Library:} Serves as a repository for software engineering and programming language research, including venues such as \textit{FSE}, \textit{TOSEM}, \textit{PLDI}, and \textit{OOPSLA}. \rev{(Search scope: \textit{Title + Abstract + Author Keywords}; 155 records).}
    \item \textbf{SpringerLink:} Offers access to journals and conference proceedings on empirical methods and data-driven software engineering. \rev{(Search scope: \textit{Keywords}; 824 records).}
    \item \textbf{ScienceDirect:} Hosts domain-specific journals, including \textit{Information and Software Technology (IST)} and \textit{Journal of Systems and Software (JSS)}. \rev{(Search scope: \textit{Title, abstract or author-specified keywords}; 26 records).}
    \item \textbf{arXiv:} Acts as the primary repository for rapid dissemination of \llm research, capturing early empirical explorations prior to formal publication. \rev{(Search scope: \textit{title + abstract}; 913 records).}
\end{itemize}

\subsection{Study Selection}
\label{sec:selection}

We established formal inclusion and exclusion criteria aligned with our research questions. \rev{Each eligible paper was treated as one unit of synthesis.} Papers satisfying all the following \textbf{Inclusion Criteria} were retained:
\begin{itemize}
    \item \textbf{Relevance to \llms and Code:} Investigates \llms or auxiliary neural approaches applied explicitly to source code generation, completion, repair, or synthesis tasks.
    \item \textbf{Focus on Quality Dimensions:} Analyzes specific quality attributes of generated code or training corpora.
    \item \textbf{Empirical Grounding:} Provides verifiable empirical evaluation, reproducible artifacts, or concrete quantitative/qualitative findings.
    \item \textbf{Publication Standards:} Published in peer-reviewed venues or available as methodologically sound preprints.
\end{itemize}

Conversely, papers meeting any of the following \textbf{Exclusion Criteria} were discarded:
\begin{itemize}
    \item \textbf{Domain Mismatch:} Focuses exclusively on natural language tasks without code-centric analysis, or addresses traditional (non-neural) code generation.
    \item \textbf{Methodological Opacity:} Lacks adequate empirical grounding, methodological transparency, or presents conceptual opinions without verifiable data.
    \item \textbf{Language and Accessibility:} Written in languages other than English or lacks full-text accessibility.
    \item \textbf{Duplication:} Represents an earlier or subsumed version of a retained extended paper.
\end{itemize}

\begin{revisionblock}
\textbf{Consolidation and De-duplication:} We consolidated the 2,298 database records and removed 184 duplicates, leaving 2,114 unique records.

\textbf{Screening Process:} Three authors independently screened the titles and abstracts of the de-duplicated records. Papers passing this phase underwent full-text review against the inclusion and exclusion criteria. Discrepancies were documented and resolved through consensus meetings. Together, these screening phases reduced the set from 2,114 records to 122 eligible papers.

\textbf{Snowballing:} We completed backward and forward snowballing in February 2026 using the 122 eligible papers as seeds~\cite{wohlin2014snowballing}. This step added 12 eligible papers, including four preprints first posted in January--February 2026, and produced 134 candidates for quality assessment.
\end{revisionblock}

\subsection{Quality Assessment}
\label{sec:quality}

\rev{We conducted a structured quality assessment of the 134 candidate papers retained after database screening and snowballing.} Each paper was independently evaluated by three reviewers against eight predefined criteria (summarized in Table~\ref{tab:quality_assessment}): (1) Relevance to Code Generation Quality, (2) Relevance to Dataset Quality, (3) Methodological Rigor, (4) \rev{Depth of Causal Analysis}, (5) Detection and Mitigation Techniques, (6) Data and Experimental Transparency, (7) Clarity and Reporting Completeness, and (8) Publication Quality.

The evaluation utilized a mixed scoring system depending on the specific criterion, with options comprising either binary (\textit{Yes}/\textit{No}) or ternary (\textit{Yes}/\textit{Partially}/\textit{No}) responses. Responses were scored as follows: \textit{Yes} = 2, \textit{Partially} = 1, and \textit{No} = 0 (maximum attainable score: $16$). \rev{We set the 10-point inclusion threshold after the initial scoring because candidates below it were insufficiently relevant to the review scope or weak on multiple methodological or reporting criteria. Across the 134 candidate papers, the three reviewers' independent pre-consensus total scores yielded Krippendorff's $\alpha = 0.717$ under the interval metric. Discrepancies were then resolved through consensus discussions, after which the threshold excluded 20 papers and retained 114 for the final synthesis. The final consensus total score for each included paper is available in the accompanying repository.}

\input{tables/quality_assessment}

\subsection{Data Extraction and Synthesis}
\label{sec:data}

\rev{We used a structured data-extraction protocol to capture information relevant to each research question.}

\paragraph{Data Extraction}
For each included paper, we extracted both bibliometric metadata and analytical findings, mapping directly to our RQs:
\begin{itemize}
    \item \textbf{Bibliographic \& Contextual Data:} Title, authors, publication year, and venue.
    \item \textbf{Code Quality Observations (RQ1):} Specific quality dimensions evaluated (e.g., correctness, security, maintainability) and the observed \codeissues.
    \item \textbf{Dataset Quality Aspects (RQ2):} Identified \dataissues (e.g., contamination, noise, imbalance, duplication).
    \item \rev{\textbf{Training-Data and Generated-Code Relationships (RQ3):} Relationships reported, analyzed, or explicitly discussed between training-data characteristics and generated-code outcomes, including the supporting passage and source-specific limitations.}
    \item \rev{\textbf{Detection \& Mitigation Interventions (RQ4 \& RQ5):} Reported diagnostic and mitigation techniques, their target issue and lifecycle stage, degree of automation, evaluation setting and outcome, required inputs or infrastructure, and reported cost, trade-offs, and limitations.}
\end{itemize}

\paragraph{Data Synthesis and Coding}
To synthesize the extracted data, we employed an iterative qualitative coding approach. The procedure consisted of three stages:
\begin{enumerate}
    \item \textbf{Open Coding:} Initial review of the papers to identify and label specific phenomena.
    \item \textbf{Axial Coding:} Grouping the initial labels into higher-level conceptual categories representing broader dimensions.
    \item \textbf{Selective Coding:} Aligning these consolidated categories with our predefined research questions (RQ1--RQ5) to establish the overarching taxonomies and \rev{relationship families}.
\end{enumerate}
\begin{revisionblock}
For the issue and technique taxonomies, coding was multi-label: a paper was assigned to every category directly addressed by its findings, so each category count is the number of papers assigned to that category and the counts may sum to more than 114.

For RQ3, we screened all included papers for within-paper relationships between an RQ2 training-data condition and an observable RQ1 outcome, then consolidated overlapping descriptions into 12 families. For the Sankey, each paper contributes one paper-level relationship assignment for each distinct combination of relationship family, RQ2 category, and RQ1 outcome; repeated passages supporting the same assignment are counted once. Figure~\ref{fig:sankey} reports the resulting coverage, and the accompanying repository provides the supporting evidence.

For RQ4 and RQ5, we compared technique families using outcomes and constraints reported in their original evaluation settings. \rev{When interpreting a result, we retained the model, language, dataset, metric, and operational conditions of its original evaluation setting}; heterogeneous effect measures were neither pooled nor used to produce an overall ranking. We classified a technique as automated when its application to an artifact did not require a person to judge each item, even if people designed its rules, prompts, or training labels.

For a coarse automation existence audit, we started from each of the eight RQ1 and thirteen RQ2 broad issue headings and reread papers reporting primary evidence until we verified one direct issue-subtype--method link for automated detection and one for automated mitigation. Each retained exemplar records the method, primary-text location, and scope limitation. The audit counts headings with at least one direct subtype-specific exemplar; it is not a count of all issue--method pairs, complete coverage of every manifestation, or effectiveness beyond the reported setting. The accompanying repository provides the 21-row crosswalk and validator.
\end{revisionblock}
Complementing the qualitative synthesis, we quantitatively aggregated the frequencies of specific issue types and mitigation strategies.

\subsection{Demographic Overview of Included Papers}
\label{sec:demographics}

The final synthesis encompasses \numstudies included papers. This section provides a demographic overview of the included literature.

\input{figures/period_distribution}

\begin{figure}[htbp]
    \centering
    \begin{minipage}{0.49\textwidth}
        \centering
        \includegraphics[width=\linewidth]{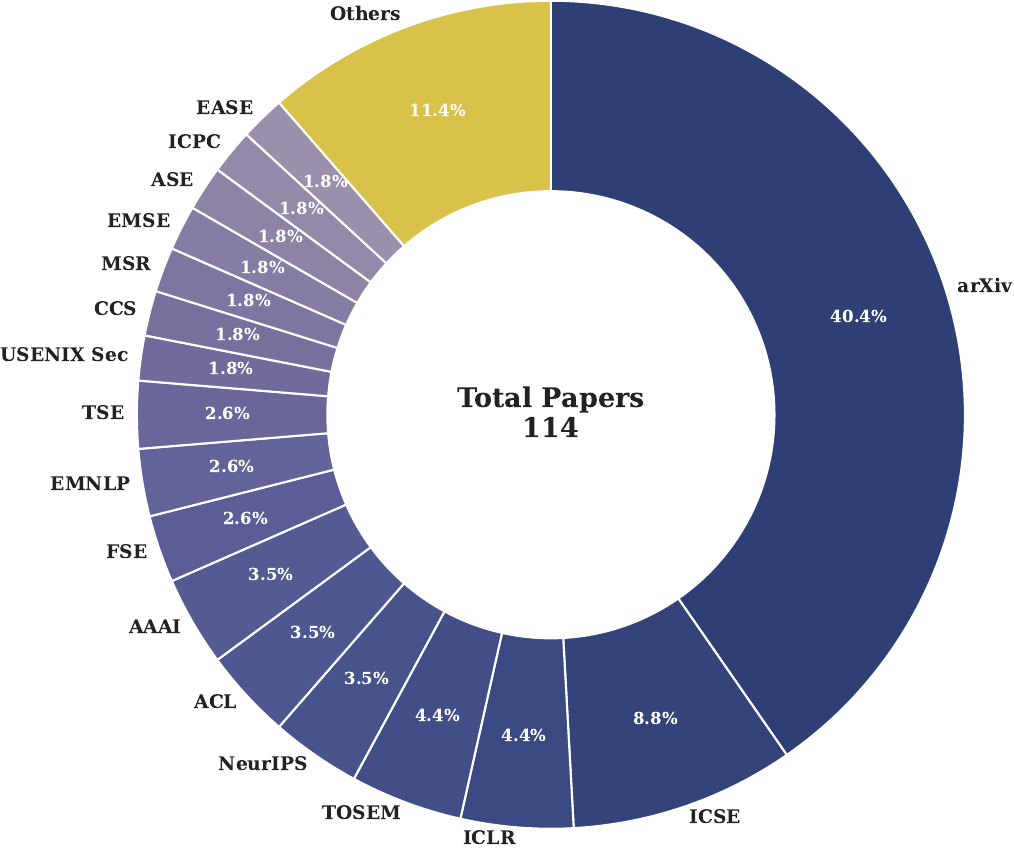}
        \caption{\rev{Publication-venue distribution of the 114 included papers. Each paper contributes once, and each slice reports its share of the 114-paper corpus; ``Others'' aggregates venues not displayed separately. The arXiv category is the largest single category (40.4\%), while the remaining papers are distributed across the separately labeled venues and ``Others.''}}
        \Description{Venue Distribution of Included Papers.}
        \label{fig:venue_donut_chart}
    \end{minipage}\hfill
    \begin{minipage}{0.49\textwidth}
        \centering
        \includegraphics[width=\linewidth]{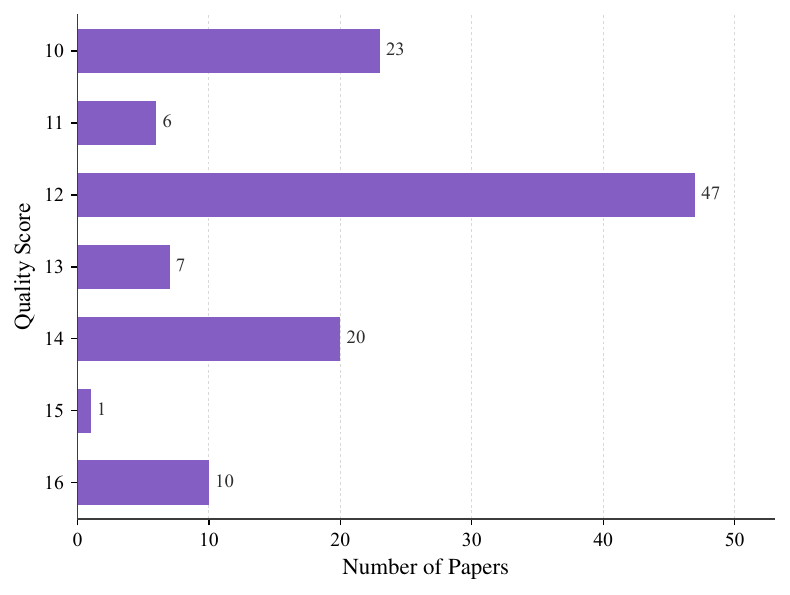}
        \caption{\rev{Quality-assessment scores of the 114 included papers. Bar length and the terminal label give the number of papers at each score from 10 to 16 on the 0--16 rubric; the counts sum to 114. Only scores at or above the inclusion threshold of 10 are shown, and 12 is the modal score (47 papers).}}
        \Description{Distribution of Included Papers by Quality Score.}
        \label{fig:quality_score_distribution}
    \end{minipage}
\end{figure}

\paragraph{Temporal Distribution}
As illustrated in Figure~\ref{fig:period_distribution}, research on \codeissues \rev{has grown rapidly}. The cumulative distribution spans from the third quarter of 2021 to the first quarter of 2026, reaching a total of \numstudies included papers. \rev{The largest increase occurred after late 2023, coinciding with the widespread release and adoption of foundational \llms.}

\paragraph{Publication Venues}
As shown in Figure~\ref{fig:venue_donut_chart}, the selected papers are distributed across a wide range of publication venues. The largest proportion is from arXiv, suggesting that this area is developing rapidly and that many results are disseminated first as preprints. In addition, the papers also appear in recognized SE venues (e.g., ICSE, FSE, ASE, TSE, TOSEM) and AI/NLP venues (e.g., ACL, NeurIPS, ICLR), indicating the interdisciplinary nature of this research topic.

\paragraph{Quality Score Distribution}
The distribution of their quality assessment scores is presented in Figure~\ref{fig:quality_score_distribution}. The scores range from 10 to 16, with the most frequent score being 12 (47 papers), followed by 10 (23 papers) and 14 (20 papers).

%% file: figures/paper_collection.tex
\begin{figure}[htbp]
    \centering
    \includegraphics[width=0.8\linewidth]{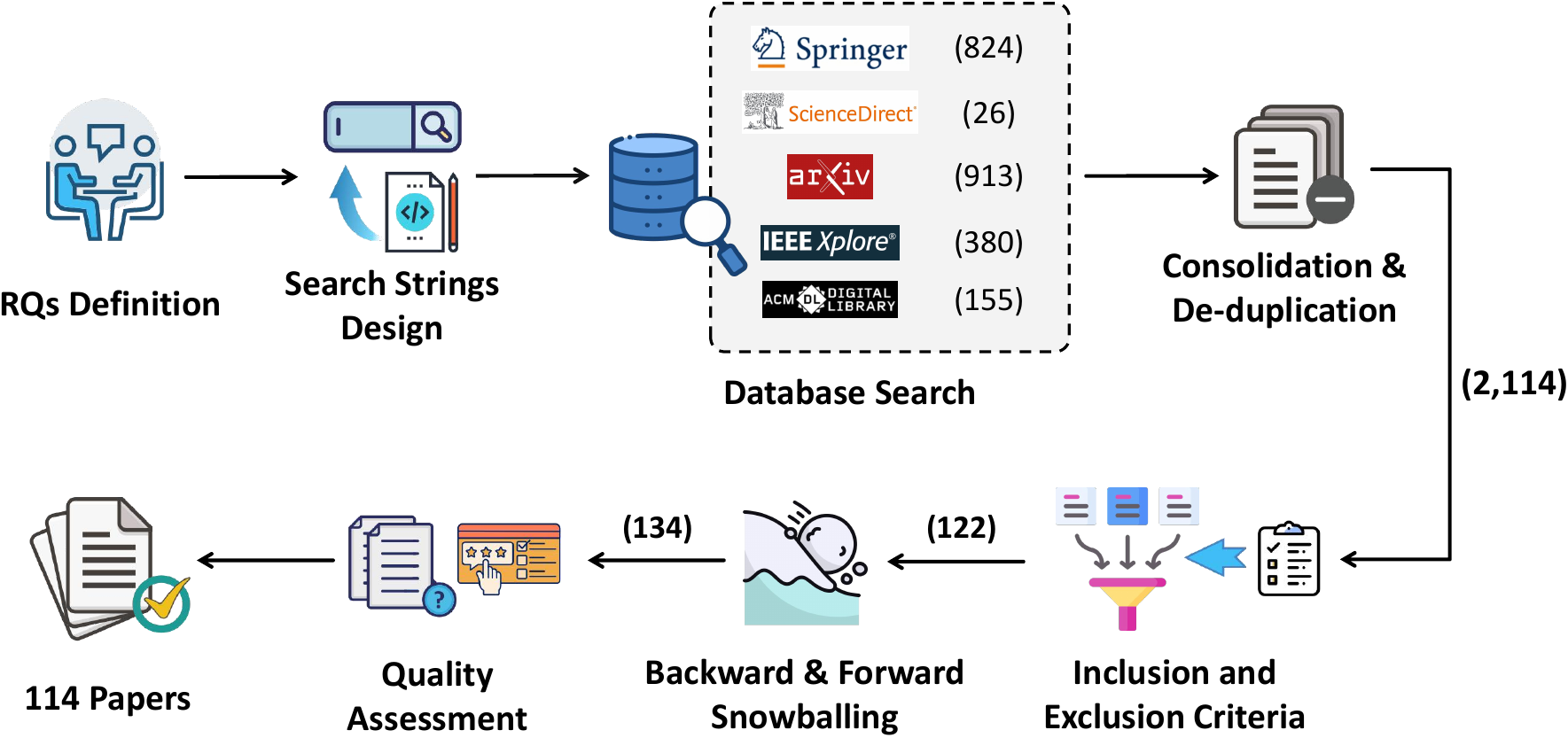}
    \caption{\rev{Study-identification and selection workflow, read from left to right along the upper row and then from right to left along the lower row. Searches of IEEE Xplore (380 records), ACM Digital Library (155), SpringerLink (824), ScienceDirect (26), and arXiv (913) returned 2,298 records. Consolidation and de-duplication retained 2,114 records; application of the inclusion and exclusion criteria retained 122 papers; backward and forward snowballing increased the set to 134; and quality assessment retained the final 114 papers.}}
    \Description{Overview of the process of paper collection and filtering.}
    \label{fig:paper_collection}
\end{figure}

%% file: tables/quality_assessment.tex
\begin{table}[!htbp]
\centering
\caption{Quality Assessment Criteria for Included Studies}
\label{tab:quality_assessment}
\renewcommand{\arraystretch}{1.4}
\begin{tabular}{p{3.5cm} p{7.5cm} c}
\toprule
\textbf{Criterion} & \textbf{Description} & \textbf{Score Options} \\
\midrule
\textbf{Relevance to Code Generation Quality} &
Whether the study investigates LLM-based code generation, completion, repair, or understanding tasks, and explicitly addresses quality aspects such as correctness, robustness, or maintainability. &
Yes / No \\

\textbf{Relevance to Dataset Quality} &
Whether the study discusses or analyzes the quality of training or fine-tuning datasets, including issues such as noise, duplication, contamination, imbalance, or outdated data. &
Yes / No \\

\textbf{Methodological Rigor} &
Whether the study clearly describes its objectives, hypotheses, research design, controlled variables, and experimental procedures, demonstrating methodological validity. &
Yes / Partially / No \\

\textbf{Depth of Causal Analysis} &
Whether the study goes beyond descriptive observations to analyze, model, or empirically verify causal relationships between data quality and generation quality. &
Yes / No \\

\textbf{Detection and Mitigation Techniques} &
Whether the study proposes concrete techniques, frameworks, or tools for detecting or mitigating code or dataset quality issues. &
Yes / No \\

\textbf{Data and Experimental Transparency} &
Whether the study releases or sufficiently documents experimental code, datasets, parameter configurations, or evaluation metrics to enable reproducibility. &
Yes / No \\

\textbf{Clarity and Reporting Completeness} &
Whether the paper presents a clear structure, comprehensive experimental reporting, and explicit discussion of limitations. &
Yes / Partially / No \\

\textbf{Publication Quality} &
Whether the study was published in a peer-reviewed venue (e.g., TSE, TOSEM, ICSE, FSE).
If it is an arXiv preprint, whether its methodological quality meets comparable standards. &
Yes / Partially / No \\
\bottomrule
\end{tabular}
\end{table}

%% file: figures/period_distribution.tex
\begin{figure}[htbp]
    \centering
    \includegraphics[width=0.7\columnwidth]{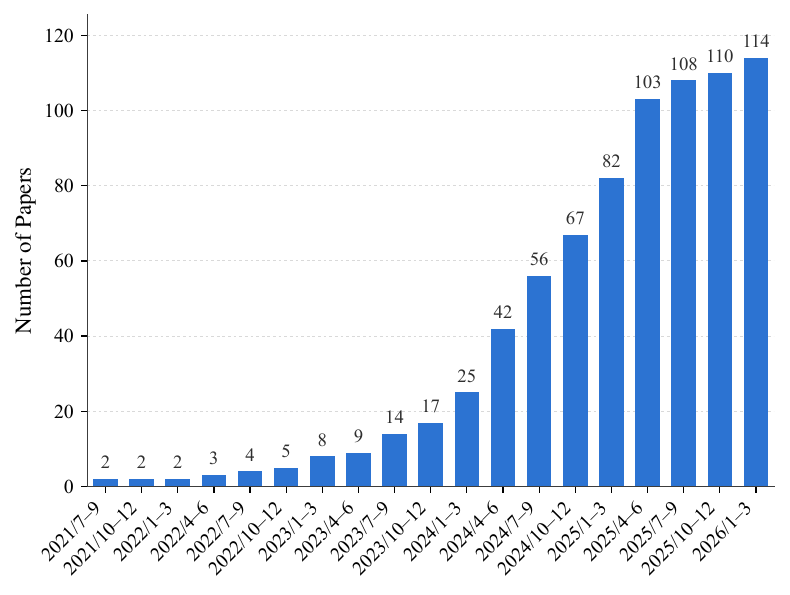}
    \caption{\rev{Cumulative temporal distribution of the 114 included papers. Each bar and its label give the cumulative number of papers included through the end of the indicated three-month publication period, increasing from 2 by 2021 Q3 to 114 by 2026 Q1; the largest increases occur during 2024--2025. Differences between adjacent bars give the number of papers added in each publication period.}}
    \Description{A bar chart showing the cumulative growth of included papers from Q3 2021 to Q1 2026, reaching a total of \numstudies papers.}
    \label{fig:period_distribution}
\end{figure}

%% file: sections/findings.tex
\section{Findings}
\label{sec:findings}

This section presents our findings organized around five research questions (RQs).
Each RQ addresses a distinct aspect of quality issues in \llm code generation and their underlying data.
The structure follows: (1) identifying \codeissues, (2) identifying \dataissues, (3) mapping between them, (4) detecting quality issues, and (5) mitigating quality issues.

\input{sections/findings/RQ1}

\input{sections/findings/RQ2}

\input{sections/findings/RQ3}

\input{sections/findings/RQ4}

\input{sections/findings/RQ5}

%% file: sections/findings/RQ1.tex
\subsection{RQ1: What \codeissues exist in \llm-generated code, and how can they be systematically categorized?}
\label{sec:rq1}

We begin by examining the spectrum of quality deficiencies exhibited in \llm-generated code. Building on the formal definition of \codeissues established in Section~\ref{sec:bgDefi} (i.e., structural, functional, or non-functional anomalies in \llm-generated code that degrade its core attributes), we synthesize findings from included papers to address the fragmentation of existing classifications. \rev{Specifically, we synthesize an evidence-informed, multi-dimensional classification of functional, non-functional, risk-related, and generation-specific quality concerns, with detailed classifications and supporting references summarized in Figure~\ref{fig:code_quality_issues_tree}.}

\input{figures/code_quality_issues_tree}

\subsubsection{Dimensions of \codeissues}
After synthesizing prior taxonomies and empirical reports, we identify \rev{eight} major and recurring dimensions that constitute our final classification of quality issues in \llm-generated code\rev{. Where a reported defect manifestation may affect multiple quality attributes, it is classified according to the immediate failure condition specified by the definitions below.}

\paragraph{\textbf{Correctness}}
Emphasizes failure to generate the intended semantics or valid execution results, with common symptoms including syntax errors, logical errors, and improper use of APIs or dependencies. These issues can manifest in different ways:

\begin{itemize}
    \item \textbf{Syntax Errors}: Occur when the generated code cannot compile or parse, often due to missing semicolons, mismatched brackets, or incomplete statements. These errors prevent the code from executing properly~\cite{10}.
    \item \textbf{Logical Errors}: Arise when the generated code produces unexpected results during runtime, even though the code compiles correctly. This could lead to runtime errors or incorrect outputs. \rev{For instance, a program may produce an incorrect result for a task-specified valid input or fail during a task-required operation under normal execution conditions, despite successful compilation~\cite{16}.}
    \item \textbf{Improper Use of APIs or Dependencies}: Happens when the generated code references non-existent functions, deprecated APIs, or uses incorrect versions of libraries. This could be due to model hallucinations or outdated training data, leading to \rev{invalid or incompatible} code. For example, a function from the deprecated \rev{\texttt{urllib2}} module in Python or a mismatch in library versions might cause the code to fail in the current environment~\cite{2}. Therefore, correctness problems are not only limited to syntax or logic errors in the code itself but are also closely tied to how the model understands and references the libraries and tools in the current software ecosystem. As training data and API versions evolve, \llms face continually changing correctness challenges, especially in long-running projects or applications spanning multiple technology stacks~\cite{24}.
\end{itemize}

\paragraph{\textbf{Security}}
\rev{Patterns that introduce weaknesses that can permit unauthorized access, disclosure, alteration, or unsafe execution.} Security issues can be further categorized into externally triggered vulnerabilities and inherent vulnerabilities, \rev{according to their primary manifestation conditions}.

\begin{itemize}
    \item \textbf{Externally Triggered Vulnerabilities}: \rev{These require the generated code to process malicious or untrusted external input for the vulnerability to be exploited, with the core issue being insufficient protection against untrusted external factors. For example, generated code may have SQL injection vulnerabilities when user input is directly concatenated into database queries without filtering; it may also render user-controlled content in a web page without appropriate output encoding, enabling cross-site scripting when another user views the page~\cite{19}.}
    \item \textbf{Inherent Vulnerabilities}: These are intrinsic design or implementation flaws that are present independently of \rev{a particular} external input. \rev{A typical example is improper encryption implementation, such as using an insecure encryption configuration or retaining an encryption key in plaintext within the application; these choices weaken the protection of encrypted data and credentials~\cite{139}.} Additionally, poor memory management and unsafe API calls are simple instances of inherent vulnerabilities, both of which introduce hidden security risks by design~\cite{53}.
\end{itemize}

\paragraph{\textbf{Compliance}}
Violations of legal, ethical, or regulatory constraints. Compliance issues typically involve generating harmful, biased, or illegal content, which may lead to legal lawsuits, ethical disputes, or privacy breaches. These issues can be further categorized into specific types:

\begin{itemize}
    \item \textbf{Malicious Toxic Code}: A typical compliance issue, distinct from security concerns, because it is specifically designed to harm others' rights and disrupt social order. Unlike security issues, which arise from \rev{vulnerabilities that may be exploited}, malicious toxic code \rev{is designed to cause damage}. For example, generated code may intentionally embed backdoors, ransomware, or other malicious software, which not only disrupt the normal functioning of the system but also violate user privacy and property rights, thereby breaching relevant laws and social norms~\cite{102}.

    \item \textbf{Privacy Issues}: \rev{These involve generated code that improperly discloses users' personal information. For example, generated code may expose user names, email addresses, or other identifying data through its handling or display, potentially violating applicable data-protection requirements~\cite{139}.}

    \item \textbf{Copyright Issues}: Occur when generated code unintentionally copies code snippets protected by copyright, violating intellectual property laws. This may happen when the model references or generates parts of code that are under copyright protection without proper attribution or permission, leading to potential legal disputes over intellectual property~\cite{42}.
\end{itemize}

These compliance issues are not just technical challenges but also involve wide-ranging legal and social responsibilities, requiring strict regulation during the code generation process.

\paragraph{\textbf{Robustness}}
\rev{The inability to handle abnormal, invalid, or out-of-range inputs, or runtime disruptions.} These issues often manifest as a lack of necessary fault tolerance and validation design in the code. For example, generated array operation code may fail to check whether the accessed index is within the array's bounds, causing an out-of-bounds error and crashing the program when an index value beyond the \rev{permitted} range is provided. Similarly, when receiving numeric data such as order amounts or ages from user input, the code may not verify whether the input is a valid number, resulting in a type conversion exception if the user accidentally enters characters or special symbols, which then interrupts the business flow~\cite{104}.

In our definition, there is a core distinction between robustness and security. Robustness focuses on the system's stability in the face of abnormal inputs, environmental fluctuations, or non-malicious disruptions, with the main goal of ensuring the program does not crash and can tolerate errors. In contrast, security is more concerned with defending against \rev{malicious attacks} (such as injection attacks or data theft), with the core objective of protecting system data and core functions from unauthorized use or damage~\cite{19}. The two concepts focus on different dimensions: ``system fault tolerance'' for robustness and \rev{``defense against security risks''} for security.

\paragraph{\textbf{Maintainability}}
Difficulty in modifying or extending code without introducing risks. This issue typically arises when code lacks clear structure or is overly complex, \rev{making modification or extension more difficult}. Common deficiencies include large, monolithic code units that handle too many responsibilities, deeply nested logic that \rev{increases the effort required to modify or extend the code}, and excessive coupling between components that reduces flexibility~\cite{68}. Additionally, code may exhibit data clumping, where multiple unrelated values are grouped together, making it difficult to adapt the system to new requirements~\cite{69}.

Weak modularization and a lack of reuse also hinder maintainability, as small changes may require significant alterations across the codebase~\cite{75}. Notably, maintainability also encompasses code reusability, which ensures that components or functions can be reused across different parts of the system without modification. Reusable components not only reduce redundancy but also make it easier to maintain and extend the system over time, as updates to one module can be reflected across all instances where it is used.

\paragraph{\textbf{Understandability}}
The ability for humans to easily comprehend the logic, intent, and structure of code. Issues often arise from poor or inconsistent naming conventions, where variable and function names fail to clearly reflect their purpose, making the code harder to follow and prone to misinterpretation~\cite{106}. Additionally, missing or misleading comments further obscure the intent of the code, leaving developers to decipher the rationale behind certain decisions or implementations. The use of ``magic numbers'', hardcoded constants without explanation, also contributes to confusion, as these values lack context and make the code less intuitive~\cite{93}.

\rev{It is important to distinguish between understandability and maintainability. Although clear and understandable code aids maintenance, maintainability itself focuses more on the code’s structure, modularity, and reusability.} Understandability, on the other hand, centers on making the code logical and transparent, ensuring that anyone can quickly grasp its functionality and intent without requiring deep expertise or context.

\paragraph{\textbf{Efficiency}}
The unnecessary consumption of computational resources in terms of time or space. These issues typically arise from inefficient computational logic design or improper resource allocation~\cite{13}.
\begin{itemize}
\item \textbf{Execution Time Efficiency}: Generated code may use suboptimal algorithms, such as employing bubble sort with a time complexity of \rev{$O(n^2)$} to process large datasets in image processing or machine learning model training, instead of using the more efficient quicksort with a time complexity of \rev{$O(n \log n)$}. This leads to a significant increase in runtime as the data size grows, severely impacting performance~\cite{29}.
\item \textbf{Memory Usage Efficiency}: There may be memory wastage due to inefficient handling of memory. For instance, in big data processing or real-time systems, the code might repeatedly create temporary objects, such as repeatedly copying large data blocks in image processing, rather than using caching or buffering. This results in frequent memory allocations and deallocations, increasing garbage collection overhead and reducing system responsiveness~\cite{45}.
\end{itemize}

\begin{revisionblock}
\paragraph{\textbf{Generation Derailment}}
Generation derailment refers to generation-specific failures in which an \llm-generated response departs from the task-directed course needed to provide a usable code deliverable consistent with the task. Such derailment can take different forms across tasks and generation settings. Representative manifestations reported in the reviewed papers include:

\begin{itemize}
    \item \textbf{Generation Misalignment}: Occurs when the response fails to deliver the task-required code artifact: it may omit, contradict, or disregard explicit instructions about task scope, implementation constraints, output format, or user feedback, or terminate before the requested code is complete. Examples include task misinterpretation, prompt-biased implementations, responses containing no code or only an empty function, prematurely terminated code generation, unrequested additions, and failure to satisfy hierarchical constraints~\cite{24,129}.

    \item \textbf{Generation Degeneration}: Occurs when the response loses its usable form through gratuitous repetition, structurally redundant code, irrelevant continuation, or unnecessary scaffolding~\cite{84,132}. Output length alone does not constitute degeneration; the defining problem is unnecessary generated content that disrupts the requested artifact.
\end{itemize}

Unlike correctness, which concerns whether code implements the intended semantics and produces valid execution results, this dimension concerns whether the response as a whole serves as the requested code artifact. Understandability and maintainability characterize the clarity and evolvability of otherwise usable code, whereas generation derailment indicates that the generated response is not an adoptable code artifact in the first place.
\end{revisionblock}

\subsubsection{Taxonomies and Classifications}
\begin{revisionblock}
Existing work includes broad surveys of \llm code generation~\cite{114}, focused taxonomies of bug patterns and generation weaknesses~\cite{24,118}, an inefficiency taxonomy~\cite{54}, and multi-dimensional quality evaluation benchmarks~\cite{93}. These papers provide valuable evidence, but their labels are framed around research developments, observed defect or inefficiency patterns, or evaluation criteria; they therefore do not provide a common taxonomy of software-quality issues in \llm-generated code.

Our eight-dimensional classification is informed by established software-quality attributes, including ISO/IEC 25010~\cite{iso25010}, and uses them as a reference rather than proposing a competing general-purpose quality model. It synthesizes functional concerns (correctness), non-functional concerns (efficiency, maintainability, understandability, and robustness), risk-related concerns (security and compliance), and generation derailment in the task-conditioned setting of \llm code generation. We treat ``hallucination,'' ``code smells,'' and ``style issues'' as cross-cutting descriptions rather than top-level dimensions because they describe generative behavior or code manifestations whose quality consequences can differ. Their specific manifestations are classified according to the affected quality attribute: non-existent APIs or attributes indicate correctness issues; task-divergent, empty, or prematurely terminated responses indicate generation derailment; structural smells that impede modification indicate maintainability issues; and naming, commenting, or formatting problems that impede comprehension indicate understandability issues.

This treatment is illustrated by empirical papers that use different vocabularies. Tambon et al.~\cite{24} distinguish ``Hallucinated Object'' and ``Wrong Attribute'' from ``Misinterpretation,'' ``Incomplete Generation,'' and ``Non-Prompted Consideration''; the former correspond to correctness, whereas the latter describe generation derailment. Abbassi et al.~\cite{54} group ``General Logic'' and ``Errors,'' ``Performance,'' ``Readability,'' and ``Maintainability'' as inefficiencies; these correspond to correctness, efficiency, understandability, and maintainability in our classification. Together, these mappings provide a common software-engineering-oriented vocabulary for comparing and synthesizing quality findings reported under heterogeneous paper-specific labels.
\end{revisionblock}
\input{figures/codeissue_date}

\begin{rqsummary}[Summary and Insights]

\begin{revisionblock}
\textbf{Research on generated code quality has expanded rapidly since 2023, particularly in 2024--2025, reflecting the accelerating development and industrial adoption of \llms for code generation (Figure~\ref{fig:CodeIssueDate}).} Early papers primarily focused on functional correctness, which remains the most frequently investigated dimension (74 papers) and continues to dominate the literature. However, the research scope has gradually broadened toward a more comprehensive view of code quality.

Security (27 papers) and maintainability (23 papers) have emerged as major concerns, reflecting growing awareness of the risks and long-term maintenance costs associated with deploying \llm-generated code in real-world systems. Meanwhile, efficiency (22 papers) and compliance (14 papers) have attracted increasing attention in 2024--2025 alongside practical deployment concerns. In contrast, understandability (13 papers), robustness (11 papers), and generation derailment (11 papers) remain comparatively underexplored. This pattern may reflect the greater difficulty of evaluating these concerns and their lower priority in existing benchmarks; generation derailment also highlights challenges specific to \llm-based code generation.

\textbf{Overall, the literature reveals a clear evolution in research priorities: from an initial focus on functional correctness toward a multi-dimensional understanding of generated code quality.} This shift suggests that the evaluation of \llm-generated code is gradually aligning with broader software quality principles, where functional correctness represents only one aspect of overall software quality. \rev{The observed diversity of issues is consistent with the evidence-informed eight-dimensional classification synthesized in this study.} Future research should further expand evaluation beyond correctness-oriented benchmarks to better capture the full spectrum of quality requirements in practical software engineering.
\end{revisionblock}

\end{rqsummary}

%% file: figures/code_quality_issues_tree.tex
\begin{figure}[htbp]
\centering
\definecolor{leafbg}{HTML}{FFFDF0}
\resizebox{\textwidth}{!}{%
\begin{forest}
for tree={
    grow'=0,
    forked edges,
    anchor=west,
    child anchor=west,
    parent anchor=east,
    font=\sffamily\footnotesize,
    inner sep=4pt,
    s sep=5pt,
    l sep=3mm,
    edge={draw=black!80},
  },
  where level=0{
    draw=purple!50,
    thick,
    rounded corners=4pt,
    fill=purple!6,
    font=\sffamily\bfseries,
    node options={
      execute at begin node={\begin{varwidth}{4.5cm}\centering},
      execute at end node={\end{varwidth}}
    }
  }{},
  where level=1{
    draw=blue!40,
    thick,
    rounded corners=4pt,
    fill=blue!6,
    node options={
      execute at begin node={\begin{varwidth}{3.5cm}\raggedright},
      execute at end node={\end{varwidth}}
    }
  }{},
  where level=2{
    draw=orange!25,
    rounded corners=3pt,
    fill=leafbg,
    node options={
      execute at begin node={\begin{varwidth}{10.5cm}\raggedright},
      execute at end node={\end{varwidth}}
    }
  }{}
  [Generated Code Quality Issues
    [Correctness
      [{LLMs Meet Library Evolution~{\scriptsize \cite{2}}, \rev{Less is More~{\scriptsize \cite{3}}}, \rev{Qwen~{\scriptsize \cite{4}}}, \rev{Qwen2~{\scriptsize \cite{5}}}, \rev{Phi-4~{\scriptsize \cite{7}}}, Copilot Evaluation~{\scriptsize \cite{10}}, HalluCode~{\scriptsize \cite{11}}, CodeHalu~{\scriptsize \cite{12}}, Mercury~{\scriptsize \cite{14}}, SStuBs~{\scriptsize \cite{16}}, Package Hallucinations~{\scriptsize \cite{17}}, HallTrigger~{\scriptsize \cite{18}}, \rev{Large Language Models for Code~{\scriptsize \cite{19}}}, The Counterfeit Conundrum~{\scriptsize \cite{23}}, Bugs in LLM-Generated Code~{\scriptsize \cite{24}}, GitHub Copilot, Amazon CodeWhisperer, ChatGPT~{\scriptsize \cite{25}}, ChatGPT Code Quality~{\scriptsize \cite{26}}, CloudAPIBench~{\scriptsize \cite{27}}, CodeMirage~{\scriptsize \cite{28}}, AutoAPIEval~{\scriptsize \cite{37}}, \rev{When Fine-Tuning LLMs Meets Data Privacy~{\scriptsize \cite{39}}}, From Effectiveness to Efficiency~{\scriptsize \cite{43}}, Software Librarian~{\scriptsize \cite{51}}, Codequal Analyzer~{\scriptsize \cite{52}}, Artificial-Intelligence Generated Code Considered Harmful~{\scriptsize \cite{53}}, Unveiling Inefficiencies in LLM-Generated Code~{\scriptsize \cite{54}}, Python Tests Quality~{\scriptsize \cite{57}}, CoQuIR~{\scriptsize \cite{58}}, REAL~{\scriptsize \cite{59}}, CIDRe~{\scriptsize \cite{61}}, Infinite-Instruct~{\scriptsize \cite{62}}, Quality In, Quality Out~{\scriptsize \cite{63}}, Security and Quality in LLM-Generated Code~{\scriptsize \cite{65}}, SwallowCode~{\scriptsize \cite{66}}, Refining ChatGPT-Generated Code~{\scriptsize \cite{69}}, ReCode~{\scriptsize \cite{74}}, Data-Efficient Fine-Tuning~{\scriptsize \cite{76}}, CRPE~{\scriptsize \cite{77}}, DeepSeek-Coder~{\scriptsize \cite{79}}, \rev{Code Pretraining~{\scriptsize \cite{81}}}, \rev{StarCoder 2 and The Stack v2~{\scriptsize \cite{82}}}, \rev{Every Sample Matters~{\scriptsize \cite{86}}}, \rev{Code Data Training Stage~{\scriptsize \cite{87}}}, \rev{WaveCoder~{\scriptsize \cite{88}}}, \rev{Brevity is the Soul of Wit~{\scriptsize \cite{89}}}, \rev{Benchmark Builders~{\scriptsize \cite{92}}}, Beyond Correctness~{\scriptsize \cite{93}}, Generated Code Diversity~{\scriptsize \cite{94}}, \rev{RedStone~{\scriptsize \cite{100}}}, \rev{Code Llama~{\scriptsize \cite{101}}}, Codex~{\scriptsize \cite{102}}, Path Planning Evaluation~{\scriptsize \cite{103}}, CODEJUDGE~{\scriptsize \cite{104}}, Synthetic Data Generation~{\scriptsize \cite{106}}, \rev{Cracks in The Stack~{\scriptsize \cite{107}}}, Unseen Horizons~{\scriptsize \cite{109}}, \rev{RTL-Breaker~{\scriptsize \cite{110}}}, \rev{MG-Verilog~{\scriptsize \cite{112}}}, Code Generation Survey~{\scriptsize \cite{114}}, DataRecipe~{\scriptsize \cite{115}}, \rev{aiXcoder-7B}~{\scriptsize \cite{117}}, Imperfect Code Generation~{\scriptsize \cite{118}}, \rev{LLM-ProS~{\scriptsize \cite{120}}}, ClassEval~{\scriptsize \cite{122}}, UCD-Training~{\scriptsize \cite{124}}, RealSec-Bench~{\scriptsize \cite{126}}, ShortCoder~{\scriptsize \cite{127}}, APIKG4SYN~{\scriptsize \cite{128}}, MultiCodeIF~{\scriptsize \cite{129}}, Adadec~{\scriptsize \cite{131}}, AllianceCoder~{\scriptsize \cite{133}}, RustEvo\textsuperscript{2}~{\scriptsize \cite{134}}, \rev{LLM Hallucinations in Practical Code Generation}~{\scriptsize \cite{137}}, AATK Benchmark~{\scriptsize \cite{139}}}]
    ]
    [Security
      [{Copilot Security~{\scriptsize \cite{9}}, Large Language Models for Code~{\scriptsize \cite{19}}, Purple Llama CYBERSECEVAL~{\scriptsize \cite{20}}, Lost at C~{\scriptsize \cite{21}}, AI Assistants Security~{\scriptsize \cite{22}}, GitHub Copilot, Amazon CodeWhisperer, ChatGPT~{\scriptsize \cite{25}}, ChatGPT Code Quality~{\scriptsize \cite{26}}, CodeMirage~{\scriptsize \cite{28}}, DeVAIC~{\scriptsize \cite{46}}, PTMs~{\scriptsize \cite{48}}, \rev{Software Librarian~{\scriptsize \cite{51}}}, Codequal Analyzer~{\scriptsize \cite{52}}, Artificial-Intelligence Generated Code Considered Harmful~{\scriptsize \cite{53}}, CoQuIR~{\scriptsize \cite{58}}, REAL~{\scriptsize \cite{59}}, Infinite-Instruct~{\scriptsize \cite{62}}, Quality In, Quality Out~{\scriptsize \cite{63}}, Security and Quality in LLM-Generated Code~{\scriptsize \cite{65}}, Codex~{\scriptsize \cite{102}}, \rev{Cracks in The Stack~{\scriptsize \cite{107}}}, Unseen Horizons~{\scriptsize \cite{109}}, \rev{RTL-Breaker~{\scriptsize \cite{110}}}, Code Generation Survey~{\scriptsize \cite{114}}, RealSec-Bench~{\scriptsize \cite{126}}, MultiCodeIF~{\scriptsize \cite{129}}, \rev{LLM Hallucinations in Practical Code Generation~{\scriptsize \cite{137}}}, AATK Benchmark~{\scriptsize \cite{139}}}]
    ]
    [Compliance
      [{AI Assistants Security~{\scriptsize \cite{22}}, DESEC~{\scriptsize \cite{38}}, When Fine-Tuning LLMs Meets Data Privacy~{\scriptsize \cite{39}}, Bias Unveiled~{\scriptsize \cite{40}}, FairCoder~{\scriptsize \cite{41}}, CodeIP~{\scriptsize \cite{42}}, CodeMI~{\scriptsize \cite{95}}, CodeCipher~{\scriptsize \cite{96}}, Code Llama~{\scriptsize \cite{101}}, Codex~{\scriptsize \cite{102}}, Synthetic Data Generation~{\scriptsize \cite{106}}, \rev{Cracks in The Stack~{\scriptsize \cite{107}}}, Code Generation Survey~{\scriptsize \cite{114}}, \rev{Asleep at the Keyboard~{\scriptsize \cite{139}}}}]
    ]
    [Robustness
      [{GitHub Copilot, Amazon CodeWhisperer, ChatGPT~{\scriptsize \cite{25}}, Bugs in LLM-Generated Code~{\scriptsize \cite{24}}, CloudAPIBench~{\scriptsize \cite{27}}, Codequal Analyzer~{\scriptsize \cite{52}}, Security and Quality in LLM-Generated Code~{\scriptsize \cite{65}}, Path Planning Evaluation~{\scriptsize \cite{103}}, CODEJUDGE~{\scriptsize \cite{104}}, Unseen Horizons~{\scriptsize \cite{109}}, MultiCodeIF~{\scriptsize \cite{129}}, Beyond Functional Correctness~{\scriptsize \cite{130}}, RobGen~{\scriptsize \cite{135}}}]
    ]
    [Maintainability
      [{Copilot Evaluation~{\scriptsize \cite{10}}, HallTrigger~{\scriptsize \cite{18}}, GitHub Copilot, Amazon CodeWhisperer, ChatGPT~{\scriptsize \cite{25}}, CodeMirage~{\scriptsize \cite{28}}, Codequal Analyzer~{\scriptsize \cite{52}}, Unveiling Inefficiencies in LLM-Generated Code~{\scriptsize \cite{54}}, Python Tests Quality~{\scriptsize \cite{57}}, CoQuIR~{\scriptsize \cite{58}}, REAL~{\scriptsize \cite{59}}, Infinite-Instruct~{\scriptsize \cite{62}}, Quality In, Quality Out~{\scriptsize \cite{63}}, Security and Quality in LLM-Generated Code~{\scriptsize \cite{65}}, ROSE~{\scriptsize \cite{68}}, Refining ChatGPT-Generated Code~{\scriptsize \cite{69}}, Seed-Coder~{\scriptsize \cite{75}}, CodeSmellEval~{\scriptsize \cite{83}}, Beyond Correctness~{\scriptsize \cite{93}}, Path Planning Evaluation~{\scriptsize \cite{103}}, CODEJUDGE~{\scriptsize \cite{104}}, Code Generation Survey~{\scriptsize \cite{114}}, DataRecipe~{\scriptsize \cite{115}}, MultiCodeIF~{\scriptsize \cite{129}}, Beyond Functional Correctness~{\scriptsize \cite{130}}}]
    ]
    [Understandability
      [{Unveiling Inefficiencies in LLM-Generated Code~{\scriptsize \cite{54}}, CIDRe~{\scriptsize \cite{61}}, Security and Quality in LLM-Generated Code~{\scriptsize \cite{65}}, Seed-Coder~{\scriptsize \cite{75}}, CRPE~{\scriptsize \cite{77}}, CodeSmellEval~{\scriptsize \cite{83}}, Beyond Correctness~{\scriptsize \cite{93}}, Synthetic Data Generation~{\scriptsize \cite{106}}, Code Generation Survey~{\scriptsize \cite{114}}, DataRecipe~{\scriptsize \cite{115}}, ShortCoder~{\scriptsize \cite{127}}, MultiCodeIF~{\scriptsize \cite{129}}, Beyond Functional Correctness~{\scriptsize \cite{130}}}]
    ]
    [Efficiency
      [{\rev{CodeHalu~{\scriptsize \cite{12}}}, EffiBench~{\scriptsize \cite{13}}, Mercury~{\scriptsize \cite{14}}, GitHub Copilot, Amazon CodeWhisperer, ChatGPT~{\scriptsize \cite{25}}, ChatGPT Code Quality~{\scriptsize \cite{26}}, LLM-Generated Code Efficiency~{\scriptsize \cite{29}}, From Effectiveness to Efficiency~{\scriptsize \cite{43}}, ENAMEL~{\scriptsize \cite{45}}, Codequal Analyzer~{\scriptsize \cite{52}}, Unveiling Inefficiencies in LLM-Generated Code~{\scriptsize \cite{54}}, CoQuIR~{\scriptsize \cite{58}}, Quality In, Quality Out~{\scriptsize \cite{63}}, Security and Quality in LLM-Generated Code~{\scriptsize \cite{65}}, Data-Efficient Fine-Tuning~{\scriptsize \cite{76}}, Beyond Correctness~{\scriptsize \cite{93}}, Path Planning Evaluation~{\scriptsize \cite{103}}, CODEJUDGE~{\scriptsize \cite{104}}, Unseen Horizons~{\scriptsize \cite{109}}, \rev{RTL-Breaker~{\scriptsize \cite{110}}}, MultiCodeIF~{\scriptsize \cite{129}}, Beyond Functional Correctness~{\scriptsize \cite{130}}, COFFE~{\scriptsize \cite{138}}}]
    ]
    [\rev{Generation Derailment}
      [{HalluCode~{\scriptsize \cite{11}}, HallTrigger~{\scriptsize \cite{18}}, Bugs in LLM-Generated Code~{\scriptsize \cite{24}}, RPG~{\scriptsize \cite{84}}, \rev{Beyond Correctness~{\scriptsize \cite{93}}}, \rev{Code Llama~{\scriptsize \cite{101}}}, Path Planning Evaluation~{\scriptsize \cite{103}}, Imperfect Code Generation~{\scriptsize \cite{118}}, DRAINCODE~{\scriptsize \cite{125}}, MultiCodeIF~{\scriptsize \cite{129}}, Code Copycat Conundrum~{\scriptsize \cite{132}}}]
    ]
  ]
\end{forest}%
}
\caption{\rev{Multi-label taxonomy of \codeissues synthesized from the 114 included papers. Read from left to right, from the root to eight dimensions and then to the relevant paper titles and citations. A paper appears in every applicable dimension, once per dimension; the total number of paper--dimension assignments may therefore exceed 114. Citation-list length measures literature coverage, with correctness having the largest citation list.}}
\Description{Taxonomy of \codeissues with corresponding literature references.}
\label{fig:code_quality_issues_tree}
\end{figure}

%% file: figures/codeissue_date.tex
\begin{figure}[htbp]
    \centering
    \includegraphics[width=0.9\linewidth]{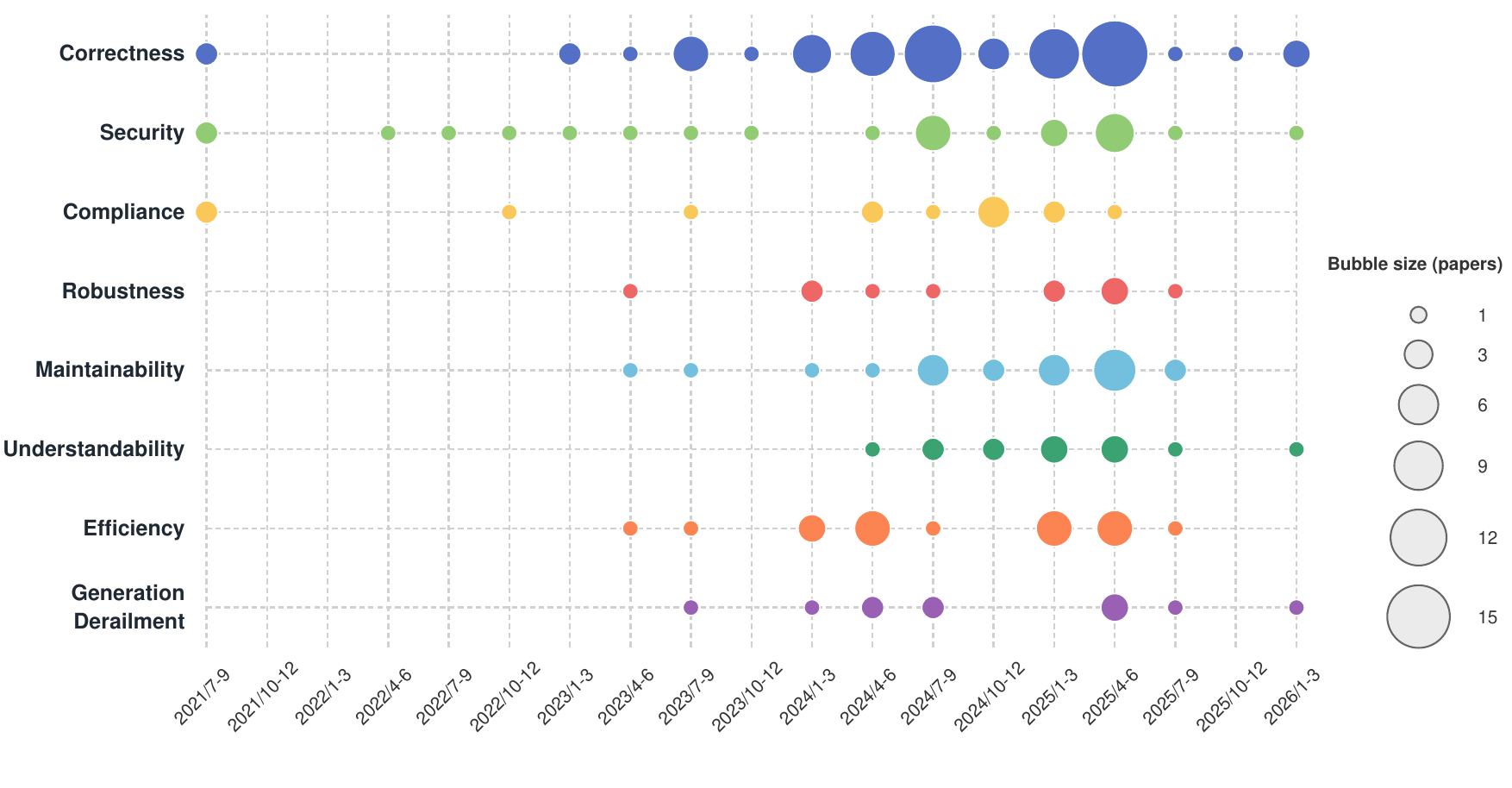}
    \caption{\rev{Three-month publication distribution of papers across the eight generated-code quality dimensions. The horizontal axis gives the publication period, the vertical axis gives the coded dimension, bubble size gives the number of included papers in that period--dimension combination (legend values: 1, 3, 6, 9, 12, and 15), and color identifies the dimension row. Because the coding is multi-label, one paper may contribute once to each applicable dimension; a missing bubble represents zero included papers for that combination. Coverage is concentrated in 2024--2025, with correctness represented most frequently.}}
    \Description{Temporal distribution of papers across eight quality dimensions, with bubble size denoting the number of papers in the corresponding period--dimension combination.}
    \label{fig:CodeIssueDate}
\end{figure}

%% file: sections/findings/RQ2.tex
\subsection{RQ2: What \dataissues exist in datasets used for \llm training, and how can they be categorized?}
\label{sec:rq2}

Next, we examine quality issues at the dataset level, \rev{which may affect} model behavior and downstream code generation. As established in Section~\ref{sec:bgDefi}, \dataissues are characterized as intrinsic flaws within pre-training and fine-tuning corpora. To provide a precise analysis, our investigation structurally divides these issues into two core dimensions: \textit{code attributes} (defects localized within individual code samples) and \textit{non-code attributes} (which encompass both textual noise and macro-level properties such as distribution, redundancy, and diversity). Specific classifications and supporting references are detailed in Figure~\ref{fig:data_quality_issues_tree}.

\input{figures/data_quality_issues_tree}

\subsubsection{Code Attribute Quality Issues}
This category investigates inherent defects within individual code samples in the training dataset. \rev{These properties correspond to conventional code-quality dimensions in RQ1. Generation Derailment applies to the complete model response and therefore appears only in the generated-code taxonomy.} The specific sub-dimensions are detailed as follows:

\paragraph{\textbf{Correctness}}
Correctness is the most fundamental quality issue in dataset code, manifested as syntax errors, logical flaws, or the failure to achieve the intended functionality. \rev{Typical examples include non-executable fragments, outdated APIs, and incorrect parameter usage~\cite{2,16,63,66}.}

\paragraph{\textbf{Security}}
Code samples in the dataset \rev{may contain inherent security vulnerabilities or high-risk programming patterns. Examples include unvalidated SQL construction, unsafe deserialization, weak cryptography, and insecure authentication or authorization mechanisms~\cite{63,107}.}

\paragraph{\textbf{Compliance}}
\rev{Compliance issues in the dataset concern code samples whose licensing, provenance, or embedded sensitive content creates legal, contractual, regulatory, or privacy risk. Examples include non-permissively licensed code, source files from repositories of uncertain provenance, and samples containing identifiable personal information or real credentials~\cite{38,95,107}.}

\paragraph{\textbf{Robustness}}
Code samples in the dataset \rev{may lack} necessary fault tolerance, making them unable to handle abnormal or out-of-range inputs, or runtime disruptions. \rev{Examples include missing boundary checks, weak input validation, and inadequate exception handling~\cite{3,130}.}

\paragraph{\textbf{Maintainability}}
Code samples in the dataset \rev{may exhibit} disorganized structures and poor extensibility, making them difficult to modify or reuse. \rev{Examples include excessive coupling, poor modularity, and code smells~\cite{63,75,115}.}

\paragraph{\textbf{Understandability}}
\rev{Code samples in the dataset may exhibit unclear logic or obscure expressions. Examples include unclear naming, unexplained constants, and comments that hinder comprehension~\cite{75,115}.}

\paragraph{\textbf{Efficiency}}
\rev{Code samples in the dataset may waste resources or contain suboptimal algorithm designs. Examples include unnecessary computation, unnecessarily high time or space complexity, and poor memory management~\cite{63,66}.}

\subsubsection{Non-Code Attribute Quality Issues}
This dimension covers all quality issues unrelated to the quality of individual code data points, focusing on defects in non-code textual data itself and macro-level attribute flaws of datasets. The specific subcategories are as follows:

\paragraph{\textbf{Compliance and Security Risks (Textual)}}
\rev{This category concerns restricted, copyrighted, or sensitive content in textual samples rather than properties of executable code.}
\begin{itemize}
    \item \rev{\textbf{Illegal and Harmful Text}: Includes content involving violence, pornography, distorted values, or speech inciting hatred or discrimination against specific groups (gender, race, age)~\cite{102,105}.}
    \item \textbf{Copyright-Infringing Text}: Copyright-protected textual data used without authorization, such as pirated technical documents, excerpts from paid tutorials, or proprietary specifications~\cite{95}.
    \item \rev{\textbf{Privacy-Leaking Text}: Text containing sensitive personal information (ID numbers, phone numbers, emails, or bank account information)~\cite{114,117}.}
\end{itemize}

\paragraph{\textbf{Distribution Imbalance Issues}}
\rev{Distribution imbalance describes disproportionate prevalence among categories represented in the corpus, including languages, domains, data types, tasks, and difficulty levels.}
\begin{itemize}
    \item \textbf{Multilingual Support Imbalance}: \rev{Mainstream languages such as Java and Python account for substantially larger shares than niche programming and non-English languages~\cite{62}.}
    \item \textbf{Domain and Resource Attention Imbalance}: \rev{Textual information is concentrated in popular technical domains and common APIs, with smaller shares devoted to niche domains and uncommon APIs. General algorithm implementations likewise account for a larger share than system architecture design or specialized-environment adaptation~\cite{6,128}.}
    \item \textbf{Data Type Proportion Imbalance}: \rev{Code and non-code data may be represented in disproportionate shares~\cite{97}.}
    \item \textbf{Data Difficulty Distribution Imbalance}: \rev{The dataset's complexity is polarized, with simple syntax examples or high-complexity algorithms accounting for larger shares than intermediate-complexity examples~\cite{77}.}
\end{itemize}

\paragraph{\textbf{Redundancy Issues}}
\rev{Redundancy consists of duplicate or near-duplicate samples that provide repetitive training signals. Synthetic data belongs to this category when it duplicates or closely resembles existing samples.}

\paragraph{\textbf{Inadequate Diversity}}
\rev{Inadequate diversity refers to relevant categories, scenarios, or knowledge for which the corpus provides no effective coverage. Examples include business logic in finance, healthcare, or industrial control, programming requirements for special operating environments such as limited computing power, and edge-case exception handling~\cite{88}.}

\paragraph{\textbf{Data Contamination Risks}}
\rev{Inadequate filtering during data collection may introduce benchmark test data (such as HumanEval or MBPP) into training sets. This can create familiarity with evaluation material and inflate measured benchmark performance, compromising assessment of generation and reasoning capabilities~\cite{82,98}.}

\paragraph{\textbf{Low-Value Data}}
\rev{Such data provides limited or misleading learning signals, including noisy, incomplete, and low-information content.}
\begin{itemize}
    \item \textbf{Meaningless Text}: Short texts containing only a URL, single words, or templated content (such as repetitive copyright notices or slogans)~\cite{98}.
    \item \textbf{Format Noise}: Residual webpage elements like navigation bars or ads, or interface text accidentally extracted from GUI elements (such as ``Login/Register'' buttons)~\cite{105}.
    \item \textbf{Low-Information-Density Text}: Spam, meaningless character sequences, invalid or malformed documents, or unparsed encoded fragments (such as Base64, hex). Texts where visible valid content is less than 20\% of the file also fall into this category~\cite{82}.
    \item \textbf{Erroneous Text}: Texts with logical fallacies, factual inaccuracies, or serious grammatical errors that mislead model understanding~\cite{7}.
    \item \textbf{Incomplete Data}: Poorly documented API references lacking parameters, explanations, or structured comments. For instance, datasets like Stack, Vault, CodeSearchNet, and MCoNaLa often lack structured annotation design, limiting the model’s ability to generate high-quality documentation or comments~\cite{61}.
\end{itemize}

\subsubsection{Taxonomies and Classifications}
\rev{Our taxonomy synthesizes \dataissues from code to textual content and from quality concerns in individual samples to properties of the training corpus as a whole.}

\rev{First, it differentiates between \textit{code attribute} and \textit{non-code attribute} issues, providing distinct source categories for the RQ3 synthesis.}
\rev{Second, the non-code attribute group separates textual-content risks from corpus-composition issues, including distribution imbalance, redundancy, inadequate diversity, contamination, and low-value data. This organization supports separate reporting and comparison of these issues.}
\rev{Third, the taxonomy represents \llm-for-code corpus concerns through subtypes such as multilingual support imbalance and data-difficulty distribution imbalance, supporting comparisons across programming languages, domains, and task settings.}

\input{figures/dataissue_date}
\begin{rqsummary}[Summary and Insights]

\textbf{\rev{Figure~\ref{fig:DataIssueDate} shows that coverage of \dataissues is concentrated in 2024--2025, mirroring the recent expansion of generated-code quality research in Section~\ref{sec:rq1}.}}

\rev{Among code attribute issues, correctness (21 papers) remains the dominant and persistent concern, reflecting its fundamental role in code-centric datasets. Security (13 papers) and maintainability (8 papers) follow, while understandability (7 papers), compliance (6 papers), efficiency (6 papers), and robustness (2 papers) receive comparatively less coverage in the reviewed literature. Their lower coverage may partly reflect the need for evidence beyond the code string itself, including licensing and provenance metadata, execution workloads, and explicit fault scenarios.}

\rev{For non-code attribute issues, distribution imbalance (29 papers) and inadequate diversity (26 papers) are the most prevalent concerns, followed by low-value data (19 papers) and redundancy (16 papers), with most coverage appearing in 2024--2025. Data contamination risks (7 papers) and textual compliance and security risks (5 papers) receive comparatively less coverage.}

\rev{These patterns suggest that LLM-for-code research and data curation should jointly assess code-attribute and non-code-attribute quality issues. Focusing on only one group leaves a distinct set of training-data quality conditions unexamined.}

\end{rqsummary}

%% file: figures/data_quality_issues_tree.tex
\begin{figure}[htbp]
\centering
\definecolor{leafbg}{HTML}{FFFFEC}
\resizebox{\textwidth}{!}{%
\begin{forest}
for tree={
    grow'=0,
    forked edges,
    anchor=west,
    child anchor=west,
    parent anchor=east,
    font=\sffamily\footnotesize,
    inner sep=4pt,
    s sep=4pt,
    l sep=3mm,
    edge={draw=black!80},
  },
  where level=0{
    draw=purple!50,
    thick,
    rounded corners=4pt,
    fill=purple!6,
    font=\sffamily\bfseries,
    node options={
      execute at begin node={\begin{varwidth}{4.5cm}\centering},
      execute at end node={\end{varwidth}}
    }
  }{},
  where level=1{
    draw=blue!40,
    thick,
    rounded corners=4pt,
    fill=blue!6,
    node options={
      execute at begin node={\begin{varwidth}{3.5cm}\raggedright},
      execute at end node={\end{varwidth}}
    }
  }{},
  where level=2{
    draw=teal!40,
    thick,
    rounded corners=4pt,
    fill=teal!4,
    node options={
      execute at begin node={\begin{varwidth}{4.0cm}\raggedright},
      execute at end node={\end{varwidth}}
    }
  }{},
  where level=3{
    draw=orange!25,
    rounded corners=3pt,
    fill=leafbg,
    node options={
      execute at begin node={\begin{varwidth}{8.5cm}\raggedright},
      execute at end node={\end{varwidth}}
    }
  }{}
  [Training Data Quality Issues
    [Code Attribute Quality Issues
      [Correctness
        [{LLMs Meet Library Evolution~{\scriptsize \cite{2}}, Less Is More~{\scriptsize \cite{3}}, SStuBs~{\scriptsize \cite{16}}, Quality In, Quality Out~{\scriptsize \cite{63}}, SwallowCode~{\scriptsize \cite{66}}, Every Sample Matters~{\scriptsize \cite{86}}, Synthetic Data Generation~{\scriptsize \cite{106}}, Cracks in The Stack~{\scriptsize \cite{107}}, RTL-Breaker~{\scriptsize \cite{110}}, MG-Verilog~{\scriptsize \cite{112}}, Code Generation Survey~{\scriptsize \cite{114}}, DataRecipe~{\scriptsize \cite{115}}, \rev{aiXcoder-7B}~{\scriptsize \cite{117}}, RustEvo\textsuperscript{2}~{\scriptsize \cite{134}}, AATK Benchmark~{\scriptsize \cite{139}}, \rev{CodeHalu~{\scriptsize \cite{12}}}, \rev{Package Hallucinations~{\scriptsize \cite{17}}}, \rev{Bugs in LLM-Generated Code~{\scriptsize \cite{24}}}, \rev{Infinite-Instruct~{\scriptsize \cite{62}}}, \rev{Code Pre-training Impact~{\scriptsize \cite{81}}}, \rev{UCD-Training~{\scriptsize \cite{124}}}}]
      ]
      [Security
        [{\rev{Copilot Security}~{\scriptsize \cite{9}}, Quality In, Quality Out~{\scriptsize \cite{63}}, StarCoder 2 and The Stack v2~{\scriptsize \cite{82}}, Cracks in The Stack~{\scriptsize \cite{107}}, RTL-Breaker~{\scriptsize \cite{110}}, AATK Benchmark~{\scriptsize \cite{139}}, \rev{Lost at C~{\scriptsize \cite{21}}}, \rev{AI Assistants Security~{\scriptsize \cite{22}}}, \rev{DeVAIC~{\scriptsize \cite{46}}}, \rev{Artificial-Intelligence Generated Code Considered Harmful~{\scriptsize \cite{53}}}, \rev{Codex~{\scriptsize \cite{102}}}, \rev{LLM Hallucinations in Practical Code Generation~{\scriptsize \cite{137}}}, \rev{Purple Llama CYBERSECEVAL~{\scriptsize \cite{20}}}}]
      ]
      [Compliance
        [{Synthetic Data Generation~{\scriptsize \cite{106}}, Cracks in The Stack~{\scriptsize \cite{107}}, Code Generation Survey~{\scriptsize \cite{114}}, Uncovering Pretraining Code in LLMs~{\scriptsize \cite{123}}, \rev{DESEC~{\scriptsize \cite{38}}}, \rev{CodeMI~{\scriptsize \cite{95}}}}]
      ]
      [Robustness
        [{Less Is More~{\scriptsize \cite{3}}, \rev{Beyond Functional Correctness}~{\scriptsize \cite{130}}}]
      ]
      [Maintainability
        [{Less Is More~{\scriptsize \cite{3}}, Quality In, Quality Out~{\scriptsize \cite{63}}, SwallowCode~{\scriptsize \cite{66}}, Seed-Coder~{\scriptsize \cite{75}}, \rev{CodeSmellEval}~{\scriptsize \cite{83}}, Synthetic Data Generation~{\scriptsize \cite{106}}, Cracks in The Stack~{\scriptsize \cite{107}}, DataRecipe~{\scriptsize \cite{115}}}]
      ]
      [Understandability
        [{Less Is More~{\scriptsize \cite{3}}, Seed-Coder~{\scriptsize \cite{75}}, Benchmark Builders~{\scriptsize \cite{92}}, Synthetic Data Generation~{\scriptsize \cite{106}}, MG-Verilog~{\scriptsize \cite{112}}, DataRecipe~{\scriptsize \cite{115}}, \rev{ShortCoder}~{\scriptsize \cite{127}}}]
      ]
      [Efficiency
        [{\rev{EffiBench}~{\scriptsize \cite{13}}, \rev{Mercury}~{\scriptsize \cite{14}}, Quality In, Quality Out~{\scriptsize \cite{63}}, SwallowCode~{\scriptsize \cite{66}}, \rev{ShortCoder}~{\scriptsize \cite{127}}, \rev{RTL-Breaker~{\scriptsize \cite{110}}}}]
      ]
    ]
    [Non-Code Attribute Quality Issues
      [Compliance and Security Risks (Textual)
        [{CodeCipher~{\scriptsize \cite{96}}, Datasets for Large Language Models~{\scriptsize \cite{105}}, \rev{aiXcoder-7B}~{\scriptsize \cite{117}}, \rev{Codex~{\scriptsize \cite{102}}}, \rev{Code Generation Survey~{\scriptsize \cite{114}}}}]
      ]
      [Distribution Imbalance Issues
        [{DataMan~{\scriptsize \cite{6}}, CIDRe~{\scriptsize \cite{61}}, Infinite-Instruct~{\scriptsize \cite{62}}, CRPE~{\scriptsize \cite{77}}, Code Pretraining~{\scriptsize \cite{81}}, Code Data Training Stage~{\scriptsize \cite{87}}, \rev{Beyond Correctness}~{\scriptsize \cite{93}}, Code Pre-Training Impact~{\scriptsize \cite{97}}, Logical Inference Pre-Training~{\scriptsize \cite{99}}, Code Llama~{\scriptsize \cite{101}}, Path Planning Evaluation~{\scriptsize \cite{103}}, Datasets for Large Language Models~{\scriptsize \cite{105}}, Synthetic Data Generation~{\scriptsize \cite{106}}, Code Generation Survey~{\scriptsize \cite{114}}, DataRecipe~{\scriptsize \cite{115}}, Training Data Extraction~{\scriptsize \cite{116}}, APIKG4SYN~{\scriptsize \cite{128}}, \rev{AATK Benchmark}~{\scriptsize \cite{139}}, \rev{Phi-4~{\scriptsize \cite{7}}}, \rev{EffiBench~{\scriptsize \cite{13}}}, \rev{CloudAPIBench~{\scriptsize \cite{27}}}, \rev{AutoAPIEval~{\scriptsize \cite{37}}}, \rev{Unveiling Inefficiencies in LLM-Generated Code~{\scriptsize \cite{54}}}, \rev{Data-Efficient Fine-Tuning~{\scriptsize \cite{76}}}, \rev{StarCoder 2 and The Stack v2~{\scriptsize \cite{82}}}, \rev{Codex~{\scriptsize \cite{102}}}, \rev{MG-Verilog~{\scriptsize \cite{112}}}, \rev{aiXcoder-7B~{\scriptsize \cite{117}}}, \rev{LLMs Meet Library Evolution~{\scriptsize \cite{2}}}}]
      ]
      [Redundancy Issues
        [{DataMan~{\scriptsize \cite{6}}, Code Pretraining~{\scriptsize \cite{81}}, StarCoder 2 and The Stack v2~{\scriptsize \cite{82}}, Repetition In Repetition Out~{\scriptsize \cite{85}}, \rev{DataComp-LM}~{\scriptsize \cite{98}}, Code Llama~{\scriptsize \cite{101}}, Datasets for Large Language Models~{\scriptsize \cite{105}}, MG-Verilog~{\scriptsize \cite{112}}, Code Generation Survey~{\scriptsize \cite{114}}, DataRecipe~{\scriptsize \cite{115}}, Training Data Extraction~{\scriptsize \cite{116}}, \rev{aiXcoder-7B}~{\scriptsize \cite{117}}, LLM-ProS~{\scriptsize \cite{120}}, \rev{CodeHalu~{\scriptsize \cite{12}}}, \rev{Data-Efficient Fine-Tuning~{\scriptsize \cite{76}}}, \rev{Code Copycat Conundrum~{\scriptsize \cite{132}}}}]
      ]
      [Inadequate Diversity
        [{WaveCoder~{\scriptsize \cite{88}}, Path Planning Evaluation~{\scriptsize \cite{103}}, Datasets for Large Language Models~{\scriptsize \cite{105}}, Synthetic Data Generation~{\scriptsize \cite{106}}, MG-Verilog~{\scriptsize \cite{112}}, Imperfect Code Generation~{\scriptsize \cite{118}}, \rev{Qwen~{\scriptsize \cite{4}}}, \rev{Qwen2~{\scriptsize \cite{5}}}, \rev{HalluCode~{\scriptsize \cite{11}}}, \rev{CodeHalu~{\scriptsize \cite{12}}}, \rev{Package Hallucinations~{\scriptsize \cite{17}}}, \rev{HallTrigger~{\scriptsize \cite{18}}}, \rev{ChatGPT Code Quality~{\scriptsize \cite{26}}}, \rev{When Fine-Tuning LLMs Meets Data Privacy~{\scriptsize \cite{39}}}, \rev{Software Librarian~{\scriptsize \cite{51}}}, \rev{Infinite-Instruct~{\scriptsize \cite{62}}}, \rev{Security and Quality in LLM-Generated Code~{\scriptsize \cite{65}}}, \rev{RedStone~{\scriptsize \cite{100}}}, \rev{Code Llama~{\scriptsize \cite{101}}}, \rev{Code Generation Survey~{\scriptsize \cite{114}}}, \rev{UCD-Training~{\scriptsize \cite{124}}}, \rev{APIKG4SYN~{\scriptsize \cite{128}}}, \rev{RustEvo\textsuperscript{2}~{\scriptsize \cite{134}}}, \rev{LLM Hallucinations in Practical Code Generation~{\scriptsize \cite{137}}}, \rev{DeVAIC~{\scriptsize \cite{46}}}, \rev{ReCode~{\scriptsize \cite{74}}}}]
      ]
      [Data Contamination Risks
        [{Phi-4~{\scriptsize \cite{7}}, StarCoder 2 and The Stack v2~{\scriptsize \cite{82}}, \rev{Beyond Correctness}~{\scriptsize \cite{93}}, DataComp-LM~{\scriptsize \cite{98}}, Datasets for Large Language Models~{\scriptsize \cite{105}}, Unseen Horizons~{\scriptsize \cite{109}}, LLM-ProS~{\scriptsize \cite{120}}}]
      ]
      [Low-Value Data
        [{\rev{Less Is More}~{\scriptsize \cite{3}}, DataMan~{\scriptsize \cite{6}}, Phi-4~{\scriptsize \cite{7}}, CIDRe~{\scriptsize \cite{61}}, StarCoder 2 and The Stack v2~{\scriptsize \cite{82}}, Every Sample Matters~{\scriptsize \cite{86}}, Brevity Is the Soul of Wit~{\scriptsize \cite{89}}, DataComp-LM~{\scriptsize \cite{98}}, Codex~{\scriptsize \cite{102}}, Datasets for Large Language Models~{\scriptsize \cite{105}}, \rev{aiXcoder-7B}~{\scriptsize \cite{117}}, \rev{Large Language Models for Code~{\scriptsize \cite{19}}}, \rev{Data-Efficient Fine-Tuning~{\scriptsize \cite{76}}}, \rev{CRPE~{\scriptsize \cite{77}}}, \rev{DeepSeek-Coder~{\scriptsize \cite{79}}}, \rev{Benchmark Builders~{\scriptsize \cite{92}}}, \rev{MG-Verilog~{\scriptsize \cite{112}}}, \rev{Code Generation Survey~{\scriptsize \cite{114}}}, \rev{DataRecipe~{\scriptsize \cite{115}}}}]
      ]
    ]
  ]
\end{forest}%
}
\caption{\rev{Multi-label taxonomy of \dataissues synthesized from the included papers. Read from left to right, the taxonomy separates code-attribute issues in individual samples from non-code-attribute issues in textual content or corpus-level properties, then lists the issue subtypes and the titles and citations of supporting papers. A paper appears under every subtype it addresses, but at most once within a subtype, so the citation lists give overlapping paper counts across subtypes. Correctness has the largest citation list among code-attribute issues, whereas distribution imbalance has the largest citation list among non-code-attribute issues.}}
\Description{Taxonomy of \dataissues with corresponding literature references.}
\label{fig:data_quality_issues_tree}
\end{figure}

%% file: figures/dataissue_date.tex
\begin{figure}[htbp]
    \centering
    \includegraphics[width=0.9\linewidth]{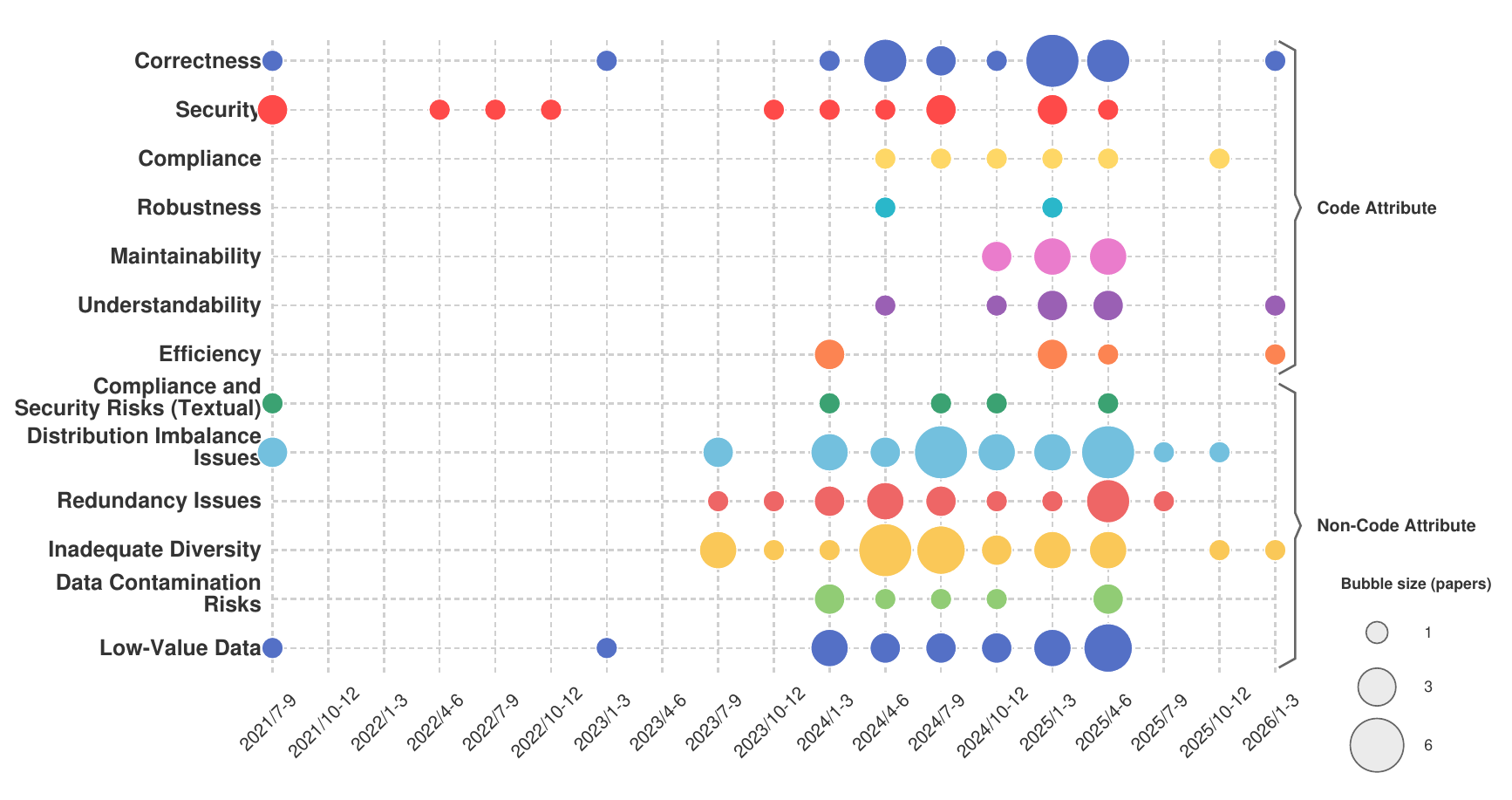}
    \caption{\rev{Three-month publication distribution of papers across training-data issue subtypes, grouped by the right-hand braces into code-attribute and non-code-attribute issues. The horizontal axis gives the publication period, the vertical axis gives the coded subtype, bubble size gives the number of included papers in that period--subtype combination (legend values: 1, 3, and 6), and color identifies the subtype row. Because the coding is multi-label, one paper may contribute once to each applicable subtype; a missing bubble represents zero included papers for that combination. Most coverage occurs in 2024--2025; correctness and distribution imbalance are the most frequently represented subtypes in their respective groups.}}
    \Description{Temporal distribution of papers across code and non-code attribute quality dimensions, with bubble size indicating paper volume per period}
    \label{fig:DataIssueDate}
\end{figure}

%% file: sections/findings/RQ3.tex
\subsection{\rev{RQ3: What relationships between \dataissues and \codeissues are reported in the literature?}}
\label{sec:rq3}

\begin{revisionblock}
RQ3 relates the RQ2 training-data taxonomy to the RQ1 generated-code taxonomy. Each mapping summarizes how included papers connect a defined training-data issue with an observable generated-code outcome.

\subsubsection{Mapping Formation and Coverage}

Comparing the reported relationships yielded 12 families. Relationships with the same training-data condition and substantive outcome were grouped together, while distinct outcomes were preserved. Thus, obsolete or otherwise defective training code belongs to \textit{Defect-Pattern Carryover}, whereas insufficient coverage of current APIs belongs to \textit{Underexposure-Associated Quality Gaps}. Comment--code inconsistencies and broader instruction--code mismatches form \textit{Supervision-Signal Quality Associations}.

Figure~\ref{fig:sankey} summarizes the literature coverage of the 12 relationship families. Section~\ref{sec:data} describes the counting rule, and the accompanying repository lists the supporting evidence.
\end{revisionblock}

\input{figures/sankey}

\begin{revisionblock}
\subsubsection{Relationship Families Centered on Code-Attribute Issues}

\paragraph{\textbf{Defect-Pattern Carryover.}}
Fourteen papers associate incorrect, non-executable, incomplete, or obsolete training examples with generated-code correctness problems (14 assignments; 13.9\%)~\cite{2,12,16,17,24,62,63,66,81,106,107,110,117,124}. The most traceable cases have a machine-checkable signature on both sides. One study measures deprecated and replacement API usages in repository proxies and then examines deprecated usages in completions~\cite{2}. A syntax-cleaning comparison reports a statistically detectable but small reduction in invalid generations, from 7.13\% to 6.68\%~\cite{63}; a controlled pre-training ablation holds architecture, token budget, non-target data, and hyperparameters fixed and finds higher correctness after removing samples that fail compilation~\cite{66}. Filtering also improves performance in one study despite using 10\% fewer examples~\cite{62}.

Together, these results support defect-specific curation: syntax and compilation checks identify training examples whose removal is associated with improved output validity, while API-version analysis connects obsolete corpus patterns with their recurrence in completions. The observed gains vary with the defect and evaluation setting.

\paragraph{\textbf{Vulnerability-Pattern Carryover.}}
Ten papers associate vulnerable training examples with generated-code security weaknesses (10 assignments; 9.9\%)~\cite{9,20,21,22,46,102,107,110,137,139}. Their designs cover corpus exposure, output behavior, and training intervention. In prompt-level comparisons, Copilot reproduces vulnerable patterns more often than fixed ones~\cite{9}. A corpus audit identifies 19,944 blobs whose newer versions fix known vulnerabilities~\cite{107}. A controlled backdoor experiment adds trigger-bearing malicious HDL examples to training and observes the implanted payload in generated code under the trigger~\cite{110}.

These findings connect security assessment at three points: locating known weaknesses in corpora, measuring their recurrence in outputs, and testing deliberately introduced training patterns. The backdoor result establishes the effect of intentional poisoning in its experimental setting; the corpus and prompt audits identify exposure and behavioral recurrence. Security-specific checks complement functional tests because vulnerable code can still compile and execute correctly. Exploitable behavior is coded here as Security; disclosure, copying, licensing, and harmful-content outcomes belong to the Compliance-centered family below.

\paragraph{\textbf{Restricted/Sensitive-Content Reproduction Risks.}}
Six papers examine code copying, license risk, secret or personal-information disclosure, and harmful or discriminatory content (eight assignments; 7.9\%)~\cite{38,53,95,102,107,114}. These form three routes to Compliance. Secret-extraction work reports 1,076 apparent real secrets~\cite{38}, while CodeMI documents concrete disclosure and licensed-code reproduction cases~\cite{95}. These results demonstrate output-side disclosure and reproduction risks. Harmful-content probes concern broader textual priors: Codex produces GPT-3-like comment biases, but the inspected GitHub fine-tuning comments do not show the same preferential co-occurrences~\cite{102}.

The family brings together exact reproduction, approximate reuse, and newly composed harmful content. Each calls for a corresponding audit: source and license comparisons for copying, targeted extraction probes for disclosure, and contextual assessment for harmful associations. The connection to a particular training source depends on provenance or membership evidence, which remains unresolved in the reported secret-extraction cases.

\paragraph{\textbf{Robustness-Pattern Associations.}}
Two papers relate fault tolerance, validation, exception handling, or boundary checking in training code to the robustness of generated programs or the correctness of generated tests (two assignments; 2.0\%)~\cite{3,130}. One finds generated code tied with or better than human ground truth on robustness in 94.2\% of annotated cases and proposes learned input-checking styles as an explanation~\cite{130}. The other removes empty exception handlers as part of a broader cleaning pipeline and reports better boundary coverage and bug detection~\cite{3}.

The two studies address complementary artifacts: fault handling in generated programs and the ability of generated tests to exercise boundaries and expose bugs. They identify input validation and exception handling as concrete data properties to examine.

\paragraph{\textbf{Structural-Quality Associations.}}
One paper associates maintainability defects in training code with generated maintainability problems (one assignment; 1.0\%)~\cite{83}. It measures code smells in generated snippets and proposes uncurated public code as an explanation. The observed result concerns output structure; the training-data connection is the study's interpretation.

\paragraph{\textbf{Clarity-Pattern Associations.}}
Two papers relate unclear naming, comments, expressions, or other comprehension defects to generated understandability problems (two assignments; 2.0\%)~\cite{106,127}. A secondary synthesis qualitatively links functionally equivalent refactoring with readability~\cite{106}. The primary experiment fine-tunes on validated original/simplified Python pairs and asks four evaluators to rate 50 outputs for comprehensibility, clarity, and relevance; functional correctness is reported separately~\cite{127}.

These approaches connect changes to training-code presentation with reader-oriented evaluation. Semantic equivalence preserves the intended behavior, while human ratings assess whether naming, comments, and expression structure communicate that behavior effectively. Brevity alone is an incomplete proxy because shortening can either remove redundancy or compress intent into unfamiliar expressions. Intrinsically obscure code belongs here; an irrelevant or inconsistent natural-language description is a supervision-signal defect.

\paragraph{\textbf{Efficiency-Pattern Carryover.}}
Four papers relate inefficient training implementations to generated efficiency problems (four assignments; 4.0\%)~\cite{13,14,110,127}. EffiBench reports that even its most efficient open model averages 2.59 times the execution time and 1.71 times the peak memory of canonical solutions, and proposes the rarity of optimal training solutions as one explanation~\cite{13}. Mercury uses fast-versus-slow solution pairs and finds that preference optimization benefits larger models more consistently than smaller ones~\cite{14}. A poisoned-HDL case reproduces a deliberately inefficient design under a trigger~\cite{110}.

The evidence highlights implementation quality and the prevalence of efficient alternatives as two data properties relevant to efficiency. Mercury supplies intervention evidence for performance-oriented preference data together with its optimization objective, with results that vary by model size. Runtime and memory comparisons among correct solutions capture the output qualities of interest.

\subsubsection{Relationship Families Centered on Non-Code and Corpus-Level Issues}

\paragraph{\textbf{Prevalence-Associated Quality Skew.}}
Seventeen papers relate prevalence or mixture imbalance to correctness, generation derailment, efficiency, or maintainability (18 assignments; 17.8\%)~\cite{2,7,13,27,37,54,76,81,82,87,93,102,103,112,117,128,139}. Controlled studies show that mixture effects depend on the evaluation setting. Phi-4 mixtures change HumanEval and MBPP differently~\cite{7}; APIKG4SYN performance rises and then falls with the single-API proportion, reaching its reported optimum at 8:2 in the tested setting~\cite{128}. MG-Verilog's balanced instruction-detail mixture is consistently near the top, with the best mixture varying across matched cells~\cite{112}.

These findings support evaluating composition against the languages, APIs, tasks, difficulty levels, and instruction formats relevant to deployment. Mixture effects vary with model capacity, schedule, post-training, and evaluation format. Observational popularity measures extend the analysis to API usage, where prevalence is associated with performance alongside differences in complexity and documentation~\cite{37,139}. Relative frequency among represented strata is prevalence skew; an absent or effectively uncovered stratum is underexposure.

\paragraph{\textbf{Redundancy-Associated Reinforcement and Degeneration.}}
Three papers relate duplicate or near-duplicate samples to correctness problems or generation derailment (three assignments; 3.0\%)~\cite{12,76,132}. CodeHalu discusses incorrect imitation of similar logic~\cite{12}, while the Code Copycat Conundrum links repetitive or truncated responses with clones and similar tests identified through a corpus analysis~\cite{132}. A selected 30--40\% instruction subset slightly outperforms full-data training; the study discusses redundancy, noise, and sample quality together in explaining the result~\cite{76}.

The reported outcomes span incorrect imitation, repetition, and truncation. Corpus analyses identify repeated patterns associated with these behaviors, while subset training provides evidence about selecting among repetitive and low-quality examples. Duplication concerns repetition within training data; contamination concerns overlap between training and evaluation material.

\paragraph{\textbf{Underexposure-Associated Quality Gaps.}}
Twenty-two papers relate insufficient task, domain, language, API, temporal, security, performance, or instruction coverage to four output dimensions (24 assignments; 23.8\%)~\cite{4,5,11,12,17,18,26,39,46,51,62,65,74,88,100,101,114,124,128,129,134,137}. Temporal and API comparisons provide concrete examples: package hallucination is about 10\% higher for recent topics~\cite{17}, RustEvo\textsuperscript{2} reports average \texttt{pass@1} of 56.1\% within prior knowledge versus 32.5\% outside training scope~\cite{134}, and equal-size API-specific fine-tuning outperforms generic instruction data in a target domain~\cite{128}.

Task-oriented studies examine coverage through different interventions. WaveCoder ablates tasks within its training mixture, changing both task coverage and volume~\cite{88}; federated bug-fix learning broadens access to private examples and reports improvements over local subsets that vary by model~\cite{39}. Across these settings, the useful unit of coverage is a concrete deployment requirement, such as a recent API, low-resource language, repository context, instruction format, or security case. This motivates a coverage matrix aligned with evaluation strata.

\paragraph{\textbf{Contamination-Associated Familiarity Differences.}}
Three papers relate verified or qualified potential training--evaluation overlap to unfamiliar-code correctness or generation derailment (three assignments; 3.0\%)~\cite{93,109,120}. In a controlled leakage experiment, task accuracy remains high while add-comments adherence falls from 47\% to 2.4\%~\cite{93}. Obfuscation and temporal comparisons find lower performance after semantics-preserving transformations and on 2024 rather than older ICPC problems~\cite{109,120}. These comparisons measure sensitivity to changed surface forms or time periods and provide indirect evidence about familiarity.

The findings show that familiarity can affect both benchmark scores and instruction following. Its measured impact varies: the GPT-4 report identifies potential overlap for 25\% of HumanEval and estimates only a small performance change on the non-contaminated subset~\cite{80}. Evaluating matched contaminated and uncontaminated tasks connects overlap with score validity, while post-cutoff and obfuscated tasks probe generalization under temporal and surface-form changes.

\paragraph{\textbf{Supervision-Signal Quality Associations.}}
Eleven papers relate noisy, low-information, malformed, irrelevant, or misaligned supervision to correctness, security, or understandability (12 assignments; 11.9\%)~\cite{3,19,61,76,77,79,89,92,112,114,115}. Manual curation of security-fix pairs removes several defect types and outperforms an uncurated set approximately 19 times as large in both security control and functional correctness~\cite{19}. A component ablation more specifically links irrelevant method--test pairs with lower compilation and coverage~\cite{3}. Rewriting descriptions improves NL-to-code metrics while changing comment authorship together with semantic consistency~\cite{92}.

The benefits depend on the task and training conditions. CIDRe filtering helps most but not all model--language cells~\cite{61}; removing long files helps short-form benchmarks under limited compute but harms long-file modeling and loses its advantage with more training~\cite{89}. DataRecipe's bundled redesign improves all evaluated models, whereas separate refactoring and semantic-correlation corrections yield heterogeneous, statistically non-significant aggregate changes~\cite{115}. Together, these results identify the signal connecting input, context, and target as a useful focus for curation. The benefit of additional data depends on its relevance and alignment with the target setting.

\subsubsection{Cross-Family Comparison and Interpretation}

The relationship graph connects two scales of data quality with generated-code outcomes. Code-attribute families usually link a defect to the corresponding output dimension, providing a concrete signature for targeted curation. Corpus-level families span several RQ1 outcomes and organize evidence about language, domain, task, and temporal coverage. This distinction supports both sample-level quality checks and corpus-level composition analysis.

Literature coverage is concentrated on Correctness, which accounts for 61 of 101 assignments (60.4\%), while Robustness, Maintainability, and Understandability together account for six. Underexposure, prevalence skew, defect carryover, and supervision-signal quality contain 68 assignments (67.3\%). Execution-based correctness measures support much of this evidence, while structural and comprehension qualities call for repository-scale or human-centered evaluation. The assignment counts describe this distribution of research attention.

Across families, controlled data interventions show that filtering, curation, and composition changes can improve quality in the evaluated settings. Corpus audits and output comparisons identify corresponding patterns, while secondary syntheses and authors' interpretations connect observations with broader data-quality explanations. For diagnosis, an obsolete API can motivate checking old examples, the prevalence of its replacement, and post-cutoff coverage; each check points to a different curation action.
\end{revisionblock}

\begin{rqsummary}[Summary and Insights]
\begin{revisionblock}

RQ3 connects generated-code quality with both the content of individual training examples and the composition of the corpus. The reviewed interventions show benefits from removing defined defects, improving supervision, and adapting data coverage or mixtures to the target task, with outcomes that vary across models and evaluation settings. Several studies obtain gains with smaller curated datasets, highlighting the value of task relevance and supervision alignment when selecting additional training examples.

The synthesis supports three complementary curation priorities: validate the quality of code examples, align input--target supervision, and match corpus coverage to deployment requirements. These priorities translate into syntax and compilation checks, validation of instruction--code pairs, and coverage assessments for target languages, APIs, and time periods, with corresponding output evaluations. Provenance and training--evaluation overlap checks support interpretation of reproduction risks and benchmark performance.
\end{revisionblock}
\end{rqsummary}
\newpage

%% file: figures/sankey.tex
\begin{figure*}[t]
    \centering
    \includegraphics[width=\textwidth]{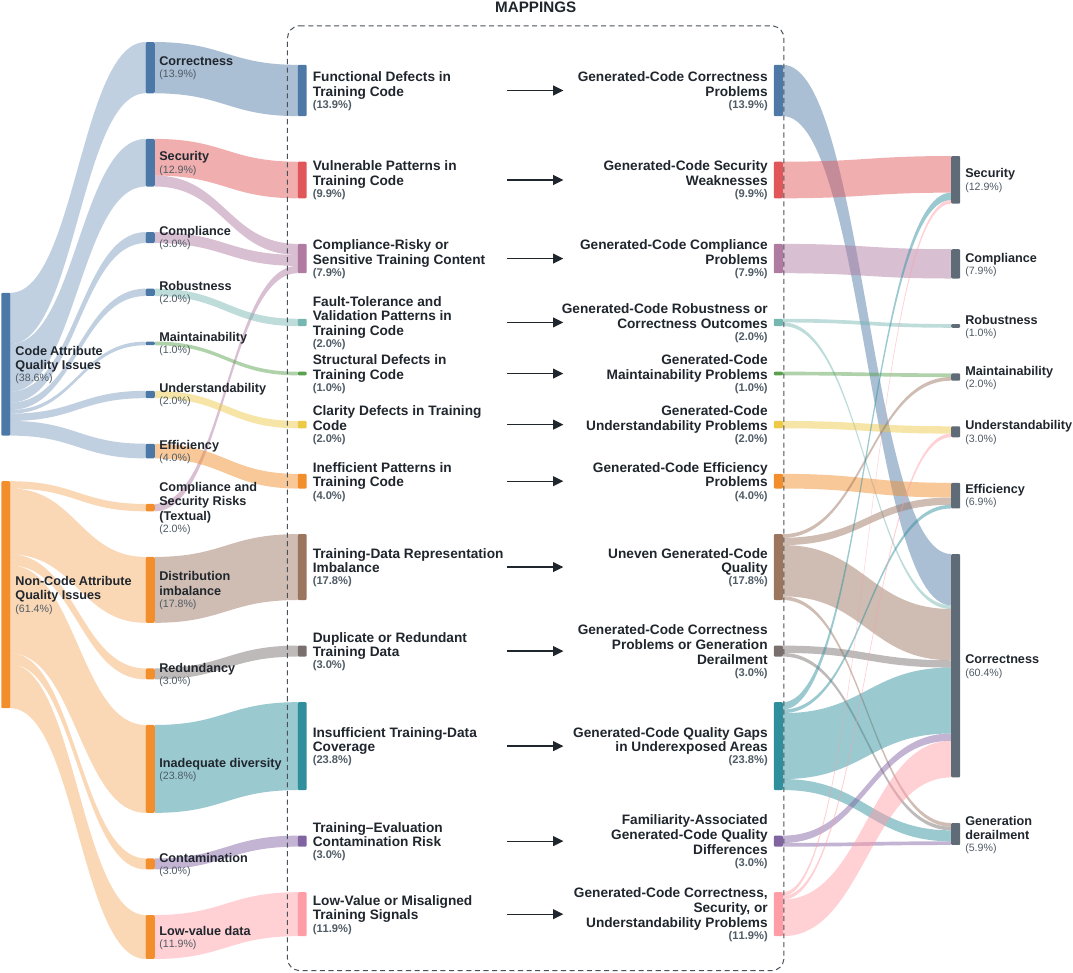}
    \caption{\rev{Relationships between training-data issues and generated-code outcomes reported in the reviewed literature. Read from left to right, all 13 RQ2 categories connect through 12 relationship families in the dashed middle region to all eight RQ1 dimensions. Band width shows the number of paper-level relationship assignments: a paper contributes one assignment for each distinct combination of relationship family, RQ2 category, and RQ1 outcome. The figure contains 101 assignments from 67 papers; percentages use 101 as their denominator and are rounded to one decimal. Width indicates literature coverage, not causal strength or effect size.}}
    \Description{A left-to-right Sankey diagram of 101 paper-level relationship assignments from 67 papers, connecting all 13 training-data issue categories through 12 relationship families to eight generated-code quality dimensions. Band widths show literature coverage.}
    \label{fig:sankey}
\end{figure*}

%% file: sections/findings/RQ4.tex
\subsection{RQ4: What techniques exist for detecting \codeissues and \dataissues?}
\label{sec:rq4}

\begin{revisionblock}
This section synthesizes detection techniques by the artifact they inspect and the information they require. For generated code, the main distinction is among behavioral evidence from execution, enumerated patterns from static checks, contextual judgments from learned evaluators, and human review. For training data, sample-level checks must be distinguished from corpus-relative audits of duplication, contamination, distribution, and provenance. Figures~\ref{fig:code_dec_tree} and~\ref{fig:data_dec_tree} retain the paper-level taxonomy, while Table~\ref{tab:detection_comparison} compares when each family is applicable, its operational advantages, and its costs and limitations. The discussion groups leaves from these multi-label taxonomies when techniques share the same evidence requirements, infrastructure, or human judgment, enabling direct comparison of applicability and limitations.
\end{revisionblock}

\subsubsection{\CodeIssue Detection}
\label{sec:rq4_code_dec}

\begin{revisionblock}
Generated-code detectors target different observables and are not interchangeable. Execution reveals behavior on exercised inputs; static rules identify encoded patterns without running the program; learned evaluators estimate semantic or contextual properties; and human review supplies project context and adjudicates consequential or ambiguous findings.
\end{revisionblock}

\input{figures/code_dec_tree}

\begin{revisionblock}
\paragraph{Execution and Resource Measurement}
Compilation, unit tests, differential testing, and profiling provide direct evidence of failures or resource use when the code is runnable and dependencies, a controlled environment, and an oracle or reference behavior are available~\cite{13,32,45,138}. This evidence remains conditional on the exercised inputs: passing a finite test set does not establish semantic equivalence or complete correctness~\cite{32}. Efficiency assessment adds further constraints. Incorrect candidates must first be excluded, workloads must be large enough to distinguish algorithms, and repeated or more stable resource measurements may be needed because wall-clock time depends on the environment~\cite{29,45,138}.

\paragraph{Static, Rule, and Provenance Analysis}
Parsers, linters, security analyzers, API maps, and overlap rules screen code without executing it and produce traceable flags for specified insecure practices, deprecated dependencies, code smells, and repetition~\cite{2,20,83,132}. Their coverage is bounded by the supported language, rule registry, threshold, version map, and available analysis context. CyberSecEval's manually checked detector, for example, attained 96\% precision and 79\% recall for the insecure coding patterns represented by its 189 rules; the remaining misses show that a configured rule set is not a complete security oracle~\cite{20}. API-aware checking similarly requires maintained version metadata and name-resolution logic~\cite{2}.

\paragraph{Model-Based and Human Assessment}
General \llm evaluators are useful when tests or fixed rules do not express intent alignment, partial correctness, robustness, or another contextual judgment. Their outputs should be treated as calibrated signals: performance varies with the evaluator, prompt, dataset, and supplied context~\cite{28,104}. Task-specific learned detectors instead require labeled, in-domain examples, as illustrated by CodeMirage's hallucination classifiers, CIDRe's structured-comment criterion, and DESEC's secret-extraction audit~\cite{28,38,61}. These methods also impose training, inference, or API costs and may shift under a new model or domain. Human review remains appropriate for unresolved or high-impact findings, but it does not scale like an automatic screen: two included audits report 66 person-hours for 58 submissions and about 108 person-hours for 333 samples, respectively~\cite{21,24}.
\end{revisionblock}

\subsubsection{\DataIssue Detection}
\label{sec:rq4_data_dec}

\begin{revisionblock}
Training-data detection separates intrinsic sample defects from corpus-relative properties. Syntax errors or encoded insecure patterns can be checked on one sample, whereas duplication, contamination, imbalance, diversity, and provenance require a comparison set, target profile, benchmark set, or repository history.
\end{revisionblock}

\input{figures/data_dec_tree}

\begin{revisionblock}
\paragraph{Execution and Static Screening}
Execution-based validation is most applicable to synthetic solutions and post-training demonstrations for which tests and a sandbox are available~\cite{5,72,73}; it does not provide a general oracle for arbitrary pre-training files. Parsers, static analyzers, metadata filters, and overlap methods operate earlier and at larger scale. One study ran 686 Python Semgrep rules over approximately 4.4 million training functions and sampled flagged results for manual validation~\cite{63}. Other pipelines use license metadata, PII detectors, MinHash or string matching, and benchmark-specific filters~\cite{7,82,117,119}. These methods depend on language-specific rules and thresholds; lexical or structural overlap does not by itself establish semantic equivalence, infringement, or an adequate target distribution.

\paragraph{Provenance, Membership, and Model-Based Screening}
Repository history and file hashes can trace the origin and versions of exact code blobs, but require large-scale version-history infrastructure and lose context after rebasing or transformation~\cite{107}. Without corpus access, CodeMI estimates whether a candidate was used for training from model probabilities and shadow models; it neither scans a corpus nor determines copyright status~\cite{95}. \llm screening and learned classifiers can score semantic quality, domain, response validity, comment consistency, or PII~\cite{6,72,73,92,117}, shifting rather than eliminating human and computational cost. DataMan, for example, used \$13,858 in super-\llm labeling for 35,700 documents; its prompt validation covered 20 documents, and its learned Qwen2-1.5B scorer achieved about 80\% average test accuracy in that study~\cite{6}.
\end{revisionblock}

\input{tables/detection_comparison}

\begin{rqsummary}[Summary and Insights]
\begin{revisionblock}
\textbf{Method choice begins with the available evidence.} Execution gives the strongest behavioral signal for exercised inputs but requires runnable artifacts and an oracle. Static and provenance checks are auditable and suitable for early, high-volume screening but cover only encoded patterns and available metadata. Learned and \llm-based assessment extends screening to contextual properties, while adding calibration, inference, and domain-shift concerns. Human review is therefore best targeted at unresolved findings and high-consequence decisions rather than used as an unbounded corpus-wide step.

\textbf{Automation shifts effort to test, rule, data, and rubric construction.} None of the families provides a general detector across all issue manifestations. The included papers do not report an automatic legal determination of copyright infringement, an intent judgment from code alone, a target-independent decision that a corpus is sufficiently diverse, or complete detection of semantic contamination and repository-level architectural quality. A layered workflow can consequently apply inexpensive deterministic checks first, targeted execution where an oracle exists, calibrated model-based triage for contextual cases, and human review for residual or high-risk decisions.
\end{revisionblock}
\end{rqsummary}

%% file: figures/code_dec_tree.tex
\begin{figure}[htbp]
\centering
\definecolor{leafbg}{HTML}{FFFDF0}
\resizebox{\textwidth}{!}{%
\begin{forest}
for tree={
    grow'=0,
    forked edges,
    anchor=west,
    child anchor=west,
    parent anchor=east,
    font=\sffamily\small,
    inner sep=5pt,
    s sep=6pt,
    l sep=3mm,
    edge={draw=black!80},
  },
  where level=0{
    draw=purple!50,
    thick,
    rounded corners=4pt,
    fill=purple!6,
    font=\sffamily\bfseries,
    node options={
      execute at begin node={\begin{varwidth}{4.5cm}\centering},
      execute at end node={\end{varwidth}}
    }
  }{},
  where level=1{
    draw=blue!40,
    thick,
    rounded corners=4pt,
    fill=blue!6,
    node options={
      execute at begin node={\begin{varwidth}{3.5cm}\raggedright},
      execute at end node={\end{varwidth}}
    }
  }{},
  where level=2{
    draw=teal!40,
    thick,
    rounded corners=4pt,
    fill=teal!4,
    node options={
      execute at begin node={\begin{varwidth}{4.0cm}\raggedright},
      execute at end node={\end{varwidth}}
    }
  }{},
  where level=3{
    draw=orange!25,
    rounded corners=3pt,
    fill=leafbg,
    node options={
      execute at begin node={\begin{varwidth}{8.5cm}\raggedright},
      execute at end node={\end{varwidth}}
    }
  }{}
  [Detection Techniques
    [Dynamic Analysis
      [Test-Based Execution Analysis
        [{Copilot Evaluation~{\scriptsize \cite{10}}, \rev{CodeHalu}~{\scriptsize \cite{12}}, Mercury~{\scriptsize \cite{14}}, HallTrigger~{\scriptsize \cite{18}}, Lost at C~{\scriptsize \cite{21}}, The Counterfeit Conundrum~{\scriptsize \cite{23}}, Bugs in LLM-Generated Code~{\scriptsize \cite{24}}, GitHub Copilot, Amazon CodeWhisperer, ChatGPT~{\scriptsize \cite{25}}, ChatGPT Code Quality~{\scriptsize \cite{26}}, LLM-Generated Code Efficiency~{\scriptsize \cite{29}}, Syntactic Robustness~{\scriptsize \cite{32}}, Bias Unveiled~{\scriptsize \cite{40}}, From Effectiveness to Efficiency~{\scriptsize \cite{43}}, ENAMEL~{\scriptsize \cite{45}}, PTMs~{\scriptsize \cite{48}}, Artificial-Intelligence Generated Code Considered Harmful~{\scriptsize \cite{53}}, Python Tests Quality~{\scriptsize \cite{57}}, REAL~{\scriptsize \cite{59}}, Quality In, Quality Out~{\scriptsize \cite{63}}, Security and Quality in LLM-Generated Code~{\scriptsize \cite{65}}, SwallowCode~{\scriptsize \cite{66}}, Refining ChatGPT-Generated Code~{\scriptsize \cite{69}}, ReCode~{\scriptsize \cite{74}}, Data-Efficient Fine-Tuning~{\scriptsize \cite{76}}, CRPE~{\scriptsize \cite{77}}, DeepSeek-Coder~{\scriptsize \cite{79}}, Beyond Correctness~{\scriptsize \cite{93}}, Codex~{\scriptsize \cite{102}}, Path Planning Evaluation~{\scriptsize \cite{103}}, Synthetic Data Generation~{\scriptsize \cite{106}}, Unseen Horizons~{\scriptsize \cite{109}}, Code Generation Survey~{\scriptsize \cite{114}}, DataRecipe~{\scriptsize \cite{115}}, ClassEval~{\scriptsize \cite{122}}, RealSec-Bench~{\scriptsize \cite{126}}, ShortCoder~{\scriptsize \cite{127}}, APIKG4SYN~{\scriptsize \cite{128}}, MultiCodeIF~{\scriptsize \cite{129}}, Beyond Functional Correctness~{\scriptsize \cite{130}}, Adadec~{\scriptsize \cite{131}}, AllianceCoder~{\scriptsize \cite{133}}, RustEvo\textsuperscript{2}~{\scriptsize \cite{134}}, COFFE~{\scriptsize \cite{138}}}]
      ]
      [\rev{Resource and Performance Measurement}
        [{LLM-Generated Code Efficiency~{\scriptsize \cite{29}}, \rev{ENAMEL}~{\scriptsize \cite{45}}, \rev{COFFE}~{\scriptsize \cite{138}}}]
      ]
    ]
    [Static Analysis
      [Rule-Based Detection
        [{LLMs Meet Library Evolution~{\scriptsize \cite{2}}, Copilot Security~{\scriptsize \cite{9}}, Copilot Evaluation~{\scriptsize \cite{10}}, SStuBs~{\scriptsize \cite{16}}, Package Hallucinations~{\scriptsize \cite{17}}, HallTrigger~{\scriptsize \cite{18}}, Large Language Models for Code~{\scriptsize \cite{19}}, Purple Llama CYBERSECEVAL~{\scriptsize \cite{20}}, Lost at C~{\scriptsize \cite{21}}, AI Assistants Security~{\scriptsize \cite{22}}, Bugs in LLM-Generated Code~{\scriptsize \cite{24}}, GitHub Copilot, Amazon CodeWhisperer, ChatGPT~{\scriptsize \cite{25}}, CloudAPIBench~{\scriptsize \cite{27}}, FairCoder~{\scriptsize \cite{41}}, DeVAIC~{\scriptsize \cite{46}}, Codequal Analyzer~{\scriptsize \cite{52}}, Artificial-Intelligence Generated Code Considered Harmful~{\scriptsize \cite{53}}, REAL~{\scriptsize \cite{59}}, Quality In, Quality Out~{\scriptsize \cite{63}}, Security and Quality in LLM-Generated Code~{\scriptsize \cite{65}}, Refining ChatGPT-Generated Code~{\scriptsize \cite{69}}, Seed-Coder~{\scriptsize \cite{75}}, CodeSmellEval~{\scriptsize \cite{83}}, RPG~{\scriptsize \cite{84}}, Path Planning Evaluation~{\scriptsize \cite{103}}, MultiCodeIF~{\scriptsize \cite{129}}, Code Copycat Conundrum~{\scriptsize \cite{132}}}]
      ]
      [Human Review and Manual Inspection
        [{Copilot Security~{\scriptsize \cite{9}}, \rev{HalluCode}~{\scriptsize \cite{11}}, HallTrigger~{\scriptsize \cite{18}}, Lost at C~{\scriptsize \cite{21}}, AI Assistants Security~{\scriptsize \cite{22}}, The Counterfeit Conundrum~{\scriptsize \cite{23}}, Bugs in LLM-Generated Code~{\scriptsize \cite{24}}, GitHub Copilot, Amazon CodeWhisperer, ChatGPT~{\scriptsize \cite{25}}, Syntactic Robustness~{\scriptsize \cite{32}}, Security and Quality in LLM-Generated Code~{\scriptsize \cite{65}}, Synthetic Data Generation~{\scriptsize \cite{106}}, Code Generation Survey~{\scriptsize \cite{114}}, Beyond Functional Correctness~{\scriptsize \cite{130}}, \rev{LLM Hallucinations in Practical Code Generation}~{\scriptsize \cite{137}}}]
      ]
    ]
    [Model-Based Detection
      [\rev{LLM-Based Evaluation}
        [{The Counterfeit Conundrum~{\scriptsize \cite{23}}, CodeMirage~{\scriptsize \cite{28}}, Unveiling Inefficiencies in LLM-Generated Code~{\scriptsize \cite{54}}, CoQuIR~{\scriptsize \cite{58}}, Seed-Coder~{\scriptsize \cite{75}}, Generated Code Diversity~{\scriptsize \cite{94}}, CODEJUDGE~{\scriptsize \cite{104}}, RealSec-Bench~{\scriptsize \cite{126}}, MultiCodeIF~{\scriptsize \cite{129}}, RobGen~{\scriptsize \cite{135}}}]
      ]
      [\rev{Task-Specific Learned Detection}
        [{\rev{CodeMirage}~{\scriptsize \cite{28}}, DESEC~{\scriptsize \cite{38}}, \rev{CIDRe}~{\scriptsize \cite{61}}}]
      ]
    ]
  ]
\end{forest}%
}
\caption{\rev{Multi-label taxonomy of techniques used to detect \codeissues. Read from left to right, the hierarchy separates execution and resource measurement, static and human analysis, and model-based assessment; it then lists papers that report each technique. A paper may appear in multiple leaves when it combines techniques. The citation lists identify papers placed in each technique; they do not establish comparative effectiveness or complete issue coverage.}}
\Description{\rev{Taxonomy of generated-code issue detection techniques, including execution, resource measurement, static rules, human review, LLM evaluation, and task-specific learned detection. Terminal boxes identify papers assigned to each technique rather than method effectiveness or complete issue coverage.}}
\label{fig:code_dec_tree}
\end{figure}

%% file: figures/data_dec_tree.tex
\begin{figure}[htbp]
\centering
\definecolor{leafbg}{HTML}{FFFDF0}
\resizebox{\textwidth}{!}{%
\begin{forest}
for tree={
    grow'=0,
    forked edges,
    anchor=west,
    child anchor=west,
    parent anchor=east,
    font=\sffamily\small,
    inner sep=5pt,
    s sep=6pt,
    l sep=3mm,
    edge={draw=black!80},
  },
  where level=0{
    draw=purple!50,
    thick,
    rounded corners=4pt,
    fill=purple!6,
    font=\sffamily\bfseries,
    node options={
      execute at begin node={\begin{varwidth}{4.5cm}\centering},
      execute at end node={\end{varwidth}}
    }
  }{},
  where level=1{
    draw=blue!40,
    thick,
    rounded corners=4pt,
    fill=blue!6,
    node options={
      execute at begin node={\begin{varwidth}{3.5cm}\raggedright},
      execute at end node={\end{varwidth}}
    }
  }{},
  where level=2{
    draw=teal!40,
    thick,
    rounded corners=4pt,
    fill=teal!4,
    node options={
      execute at begin node={\begin{varwidth}{4.0cm}\raggedright},
      execute at end node={\end{varwidth}}
    }
  }{},
  where level=3{
    draw=orange!25,
    rounded corners=3pt,
    fill=leafbg,
    node options={
      execute at begin node={\begin{varwidth}{8.5cm}\raggedright},
      execute at end node={\end{varwidth}}
    }
  }{}
  [Training Data Quality Issue Detection
    [Dynamic Analysis
      [Execution-Based Validation
        [{Qwen2~{\scriptsize \cite{5}}, \rev{TeleChat2}~{\scriptsize \cite{72}}, Kimi K2~{\scriptsize \cite{73}}}]
      ]
    ]
    [\rev{Static and Corpus Audit}
      [\rev{Syntax, Rule, Metadata, and Overlap Checks}
        [{Less Is More~{\scriptsize \cite{3}}, Qwen~{\scriptsize \cite{4}}, Qwen2~{\scriptsize \cite{5}}, Quality In, Quality Out~{\scriptsize \cite{63}}, SwallowCode~{\scriptsize \cite{66}}, \rev{TeleChat2}~{\scriptsize \cite{72}}, Data-Efficient Fine-Tuning~{\scriptsize \cite{76}}, CRPE~{\scriptsize \cite{77}}, DeepSeek-Coder~{\scriptsize \cite{79}}, StarCoder 2 and The Stack v2~{\scriptsize \cite{82}}, Repetition In Repetition Out~{\scriptsize \cite{85}}, Every Sample Matters~{\scriptsize \cite{86}}, WaveCoder~{\scriptsize \cite{88}}, Brevity Is the Soul of Wit~{\scriptsize \cite{89}}, DataComp-LM~{\scriptsize \cite{98}}, Codex~{\scriptsize \cite{102}}, Datasets for Large Language Models~{\scriptsize \cite{105}}, Cracks in The Stack~{\scriptsize \cite{107}}, MG-Verilog~{\scriptsize \cite{112}}, DataRecipe~{\scriptsize \cite{115}}, Training Data Extraction~{\scriptsize \cite{116}}, \rev{aiXcoder-7B}~{\scriptsize \cite{117}}, Inter-Dataset Code Duplication~{\scriptsize \cite{119}}, LLM-ProS~{\scriptsize \cite{120}}, AATK Benchmark~{\scriptsize \cite{139}}}]
      ]
      [Human-in-the-Loop Review
        [{Qwen~{\scriptsize \cite{4}}, DataMan~{\scriptsize \cite{6}}, \rev{TeleChat2}~{\scriptsize \cite{72}}, StarCoder 2 and The Stack v2~{\scriptsize \cite{82}}, Datasets for Large Language Models~{\scriptsize \cite{105}}, Synthetic Data Generation~{\scriptsize \cite{106}}, Imperfect Code Generation~{\scriptsize \cite{118}}, AATK Benchmark~{\scriptsize \cite{139}}}]
      ]
      [\rev{Provenance and Version-History Audit}
        [{Cracks in The Stack~{\scriptsize \cite{107}}}]
      ]
      [\rev{Model-Membership and Attribution Audit}
        [{CodeMI~{\scriptsize \cite{95}}, \rev{Uncovering Pretraining Code in LLMs}~{\scriptsize \cite{123}}}]
      ]
    ]
    [Model-Based Detection
      [LLM-Based Dataset Screening
        [{Qwen~{\scriptsize \cite{4}}, Qwen2~{\scriptsize \cite{5}}, DataMan~{\scriptsize \cite{6}}, SwallowCode~{\scriptsize \cite{66}}, Qwen3~{\scriptsize \cite{70}}, Qwen2.5~{\scriptsize \cite{71}}, \rev{TeleChat2}~{\scriptsize \cite{72}}, Kimi K2~{\scriptsize \cite{73}}, DeepSeek-Coder~{\scriptsize \cite{79}}, WaveCoder~{\scriptsize \cite{88}}, Benchmark Builders~{\scriptsize \cite{92}}, Datasets for Large Language Models~{\scriptsize \cite{105}}, DataRecipe~{\scriptsize \cite{115}}}]
      ]
      [\rev{Task-Specific Learned Classifiers}
        [{\rev{Less Is More}~{\scriptsize \cite{3}}, \rev{DataMan}~{\scriptsize \cite{6}}, Benchmark Builders~{\scriptsize \cite{92}}, \rev{DataComp-LM}~{\scriptsize \cite{98}}, \rev{aiXcoder-7B}~{\scriptsize \cite{117}}}]
      ]
    ]
  ]
\end{forest}%
}
\caption{\rev{Multi-label taxonomy of techniques used to detect \dataissues, with papers reporting each technique. Papers may appear in multiple leaves; citations identify paper placements, not comparative effectiveness or complete issue coverage.}}
\Description{\rev{Taxonomy of training-data issue detection techniques, including execution validation, syntax and metadata checks, human review, provenance and model audits, LLM screening, and task-specific learned classifiers. Terminal boxes identify papers assigned to each technique rather than method effectiveness or complete issue coverage.}}
\label{fig:data_dec_tree}
\end{figure}

%% file: tables/detection_comparison.tex
\begin{table*}[!b]
\centering
\caption{\rev{Comparison of detection families found in the included papers and author-synthesized applicability conditions. Reported outcomes remain tied to the original languages, datasets, models, rules, and evaluation settings; the table does not rank families across heterogeneous evaluations.}}
\label{tab:detection_comparison}
\begin{revisionblock}
\small
\renewcommand{\arraystretch}{1.15}
\setlength{\tabcolsep}{3pt}
\begin{tabular}{@{}>{\raggedright\arraybackslash}p{2.3cm} >{\raggedright\arraybackslash}p{3.7cm} >{\raggedright\arraybackslash}p{3.8cm} >{\raggedright\arraybackslash}p{4.65cm}@{}}
\toprule
\textbf{Detection family} & \textbf{Applicable when} & \textbf{Operational advantage} & \textbf{Cost and evidence limitation} \\
\midrule
\textbf{Execution and resource measurement} &
Generated code or data samples can be built or executed, and tests, a reference behavior, or a representative workload is available; targets include correctness, robustness, and resource use. &
Compilation, differential or benchmark tests, and execution-time, instruction-count, or memory measurements provide direct behavioral or resource signals~\cite{13,32,45,138}. &
Requires executable dependencies, a controlled environment, test inputs, and an oracle or reference; efficiency evaluation may require stress inputs and repeated or standardized measurements~\cite{13,45,138}. Finite differential tests can expose non-equivalence but cannot prove equivalence~\cite{32}. \\

\textbf{Static, rule, metadata, and overlap analysis} &
A parser, rule registry, API/version map, provenance record, or comparison set exists; targets include specified insecure practices, obsolete APIs, code smells, licensing risk, duplication, and lexical contamination. &
Rule- and metadata-based checks can flag artifacts without executing them; the included papers apply such checks to both training corpora and generated outputs~\cite{2,20,63,82,107,119}. &
Coverage is bounded by supported languages, rules, thresholds, metadata, and comparison sets. CyberSecEval's manually checked detector attained 96\% precision and 79\% recall for its covered insecure patterns, showing both useful automation and misses~\cite{20}. \\

\textbf{General \llm evaluation} &
The target requires intent, context, partial-correctness, robustness, or qualitative judgment, and a rubric, prompt, and capable evaluator can be supplied. &
Can assess artifacts for which executable or fixed-rule oracles are unavailable and can triage findings from other tools~\cite{28,54,104,126}. &
Outputs vary with evaluator, prompt, order, context, and dataset; calibration or human validation is needed. Repeated API calls or multi-GPU local inference can constrain scale, and some studies consequently evaluate only samples or one judge run~\cite{54,94,104}. \\

\textbf{Task-specific learned scorer} &
Labeled in-domain examples or a learnable proxy signal exists, such as coverage, hallucination class, comment quality, domain, or PII. &
After training, a fixed scorer can batch-process artifacts and preserve a stable version for a particular pipeline~\cite{3,6,28,61,92,117}. &
Label construction, training, and domain shift limit transfer. DataMan's 35,700-document labeling set cost \$13,858; its human prompt check covered 20 documents, and the trained scorer achieved about 80\% average test accuracy in its general-text setting~\cite{6}. \\

\textbf{Human or hybrid review} &
Study labels or automated flags require semantic or security judgment. In the included evidence, this role is primarily research annotation or validation rather than an evaluated repository gate. &
Reviewers recategorize ambiguous security outputs, construct bug taxonomies, and check selected data after automatic filters~\cite{9,21,24,72,82}. &
Expert effort limits throughput and introduces disagreement. Two code audits report 66 person-hours for 58 submissions and about 108 person-hours for 333 samples; corpus-wide error estimates require a documented sampling design and uncertainty~\cite{21,24}. \\
\bottomrule
\end{tabular}
\end{revisionblock}
\Description{\rev{Comparison of execution, static and metadata analysis, LLM evaluation, task-specific learned scoring, and human or hybrid review by prerequisites, advantages, costs, and evidence limitations.}}
\end{table*}

%% file: sections/findings/RQ5.tex
\subsection{RQ5: What mitigation strategies have been proposed to address these quality issues?}
\label{sec:rq5}

\begin{revisionblock}
\rev{The reviewed literature reports interventions at three points: altering training or fine-tuning data, updating model behavior, and controlling or repairing a particular generation.} These points differ in deployment cost and scope. Data and model interventions require another training stage but can affect later requests, whereas generation-time controls can be applied to a fixed model but add context, decoding, tool, or repeated-inference cost. Figure~\ref{fig:code_mit_tree} organizes generated-code strategies by lifecycle point, Figure~\ref{fig:data_mit_tree} details training-data strategies, and Table~\ref{tab:mitigation_comparison} compares their conditions, reported outcomes, costs, and limitations. The discussion follows the generated-code lifecycle at a coarse level and groups the finer-grained data-mitigation leaves by the operation they perform---removal, transformation, selection, or expansion---to compare shared prerequisites and trade-offs.
\end{revisionblock}

\subsubsection{\CodeIssue Mitigation}
\label{sec:rq5_code_mit}

\begin{revisionblock}
Generated-code mitigation includes preventive changes to data or model behavior and candidate-specific controls during or after inference. The target property and available oracle determine which intervention is applicable; a method that reduces one measured issue can leave another unchanged or worse.
\end{revisionblock}

\input{figures/code_mit_tree}

\begin{revisionblock}
\paragraph{Data-Level Mitigation}
Data-level interventions improve training or fine-tuning material before a model produces code. Their detailed taxonomy appears in Figure~\ref{fig:data_mit_tree} and Section~\ref{sec:rq5_data_mit}. These interventions are indirect with respect to a later generated artifact: their effect must be established by retraining and evaluating the resulting model under held-out conditions.

\paragraph{Model Adaptation and Training-Time Control}
Fine-tuning can specialize a model for a defined target, but its effect need not transfer to other qualities or tasks. In the package-hallucination study, supervised fine-tuning lowered the measured hallucination rate for two open models while HumanEval pass@1 also declined in that experimental setting~\cite{17}. SVEN instead learned continuous security prefixes from approximately 1,600 curated examples across selected languages and CWEs; its reported security gains remained bounded by those detectors and task settings~\cite{19}. ShortCoder used Python rewrite pairs and LoRA to reduce generated tokens, while its runtime claim concerns generation latency rather than the algorithmic running time of the produced program~\cite{127}.

Reward-guided training requires executable or static feedback that represents the intended property. REAL combines unit-test rewards with analyzers for selected security and maintainability defects~\cite{59}; optimizing its quality signal alone produced trivial non-functional outputs, whereas a combined reward reduced that proxy failure. ReCode uses format, syntax, and edit-similarity rewards for API migration rather than a general correctness reward~\cite{74}. These studies show why a reward must be interpreted together with its tests and analyzer coverage. A narrower training-time control, Repetition Dropout, masks attention to repeated n-grams; it targets repetition specifically and can trade lower repetition against higher perplexity~\cite{85}.

\paragraph{Generation-Time Control and Repair}
Prompt constraints have low deployment overhead but their effect varies by model and task; some reviewed prompt variants improved a target measure while others reduced correctness~\cite{2,32,130}. CodeCipher is a distinct input transformation: it obfuscates source code sent to a remote model and trades privacy against utility, but it does not mitigate privacy, licensing, or malicious content in the generated output~\cite{96}.

Retrieval can supply current package, API, or repository information without changing model weights~\cite{17,27,133,134}. It requires a maintained corpus, relevant retrieval, and context budget. CloudAPIBench illustrates the trade-off: documentation improved low-frequency API use, while a suboptimal retriever reduced high-frequency API accuracy by 39.02 percentage points; selective triggering reduced that interference in the study setting~\cite{27}.

White-box decoding can apply grammar, uncertainty, or learned-checker signals at each step. Examples target repetition, a correctness--latency operating point, missing condition checks, or provenance watermarks~\cite{42,84,131,135}. Such controls require logits and sometimes a language grammar or trained checker. RobGen's Java evaluation reported improved robustness proxies but 33.4\%--197.9\% time overhead across configurations and inconsistent behavior for simpler prompt or insertion baselines~\cite{135}.

After generation, diagnostics can guide another model call or a deterministic repair. Static and runtime feedback supported partial repair in Java and Python, but vague feedback could be less effective, fixes could introduce new issues, and gains tended to stabilize after early rounds~\cite{69}. MultiCodeIF similarly reports increasing constraint satisfaction over four feedback rounds for two models on its synthetic benchmark~\cite{129}. Deterministic transformations are cheaper when a defect has a known form: API replacement and DeRep address mapped obsolete calls and identified repetition patterns, respectively, but do not generalize beyond their maintained mappings and rules~\cite{2,132}.
\end{revisionblock}

\subsubsection{\DataIssue Mitigation}
\label{sec:rq5_data_mit}

\begin{revisionblock}
Training-data mitigation comprises removal, transformation, selection, and expansion. Removal is appropriate when a sample violates a reliable rule; transformation is useful when recoverable information should be retained; selection changes a corpus mixture relative to a stated deployment goal; and expansion adds coverage but also requires provenance and validation.
\end{revisionblock}

\input{figures/data_mit_tree}

\begin{revisionblock}
\paragraph{Validation, Filtering, and Sanitization}
Compilers, tests, parsers, linters, security analyzers, license tools, PII models, and duplicate or contamination matching can remove samples with detectable defects before training~\cite{3,63,82,117}. These checks are automatic at application time but require test or rule construction, threshold selection, infrastructure, and exception handling. In two Java unit-test datasets, CleanTest combined syntax, relevance, and learned coverage filters; average fine-tuning time fell by 25.45\% and 10.00\%, and most aggregate outcomes improved, but individual results included a 1.26\% syntactic-correctness decline for CodeBERT and a 10.61\% compilation decline for StarCoder~\cite{3}. Its rules and evaluation were Java-specific and the authors identify possible public-repository overlap. The Stack v2 demonstrates a broader production-scale pipeline with licensing, deduplication, PII, malware, benchmark, and opt-out controls, but it does not isolate the downstream effect of each filter~\cite{82}.

\paragraph{Scoring, Rewriting, and Transformation}
Learned and \llm-based scorers retain or sample data according to a quality, domain, relevance, coverage, or comment criterion~\cite{3,6,61,98}. Their threshold is a policy choice, and their labels and domain determine what they preserve. Rewriting instead attempts to repair a useful sample. SwallowCode applies Python syntax and Pylint filtering followed by style-guided and semantic rewriting~\cite{66}. In its 50-billion-token continual-pretraining setting, rewriting used 22\% more computation than \llm scoring and produced larger HumanEval-family gains, while automated renaming caused an approximately 10-point MBPP decline and SCOR was not ablated independently of the preceding stage~\cite{66}. DataRecipe likewise found that rule/tool transformations and \llm intent--code correction varied by model and benchmark, with no significant overall improvement in its reported aggregate tests~\cite{115}. Transformation therefore needs interface-preservation checks and downstream reevaluation rather than an assumption of monotonic improvement.

\paragraph{Mixture Design and Dataset Expansion}
Quality-aware sampling and mixture design use labels, domains, languages, task types, or model capacity to specify a training distribution~\cite{4,6,70,72,81,101}. DataMan provides a controlled general-text example rather than code-generation evidence, while Qwen and Code Llama describe complete training pipelines whose individual mixture choices are not independently evaluated. A target mixture must consequently be justified against intended deployment, not equated with a universal balance or fairness guarantee.

Curated-source and synthetic expansion add languages, tasks, domains, solutions, or reasoning trajectories~\cite{5,7,71,74,77,82,101,124,128}. External origin does not by itself establish quality; new material still requires licensing, provenance, deduplication, and sample validation. Synthetic pipelines likewise inherit generator and oracle biases and incur generation and verification cost. Their evidence is task-specific, so improvements from different models, corpus sizes, and benchmarks are not pooled.
\end{revisionblock}

\input{tables/mitigation_comparison}

\begin{revisionblock}
\paragraph{Human Roles}
Automation changes where expert work occurs. Included pipelines use people to define rules and labels, review file types or data subsets, verify API migrations, select target mixtures and thresholds, and adjudicate uncertain security or licensing cases~\cite{6,72,74,82}. This role supports calibration and exceptions; review of a sample does not certify every unreviewed item.
\end{revisionblock}

\begin{revisionblock}
\paragraph{Broad-Heading Automation Existence Audit}
We manually checked each of the eight RQ1 and thirteen RQ2 broad headings for direct primary empirical evidence of automated detection and mitigation. Across both artifact layers, every heading had at least one subtype with a direct automated detection exemplar and at least one---though not necessarily the same---subtype with a direct automated mitigation exemplar. These 8/8 generated-code and 13/13 training-data results establish only the existence of automation evidence at the broad-heading level; they do not indicate complete subtype coverage, paired detection and mitigation, deployment readiness, or effectiveness beyond the reported setting. Generated-code Compliance, for example, is represented by a white-box secret-leakage audit for detection and a pre-embedded provenance watermark for mitigation; Understandability is represented by structured-comment quality, Generation Derailment by repetition, and training-data Robustness by empty exception handling in Java test data~\cite{3,38,42,61,132}. The repository crosswalk records a method, primary-text location, and scope limitation for every heading. The audit therefore shows that no whole heading lacks all automatic evidence, while the subtype and deployment gaps above remain unresolved.
\end{revisionblock}

\begin{rqsummary}[Summary and Insights]
\begin{revisionblock}
\textbf{No mitigation family dominates across quality issues.} Model updates can amortize a behavior change across future generations but require data, training, and regression evaluation. Prompting and retrieval are accessible for fixed or closed models but depend on instruction following and relevant context. White-box decoding offers finer control at additional latency and integration cost. Feedback-guided repair is most actionable when tools return a localized diagnostic, while deterministic repair is limited to encoded patterns.

\textbf{Multiple signals reduce, but do not eliminate, proxy conflict.} Tests miss unexercised behavior, static analyzers miss unencoded defects, retrieval can inject irrelevant context, and data transformations can break interfaces or shift distributions. The included results therefore support rechecking all required properties after each intervention and retaining a human decision path for legal, security, architectural, and other high-consequence ambiguity.

\textbf{The clearest gaps are at issue-manifestation and deployment levels.} \rev{The included literature reports} narrow automatic interventions for broad categories such as robustness, compliance, and generation derailment, but does not establish complete coverage of their subtypes. Direct evaluations remain sparse for real runtime or memory optimization of generated programs, output-side copyright and privacy leakage, semantic contamination, downstream risk reduction after license or PII filtering, and the isolated contribution of diversity or mixture changes. End-to-end evaluations in maintained repositories and CI/CD pipelines are also uncommon.
\end{revisionblock}
\end{rqsummary}

%% file: figures/code_mit_tree.tex
\begin{figure}[htbp]
\centering
\definecolor{leafbg}{HTML}{FFFDF0}
\resizebox{\textwidth}{!}{%
\begin{forest}
for tree={
    grow'=0,
    forked edges,
    anchor=west,
    child anchor=west,
    parent anchor=east,
    font=\sffamily\small,
    inner sep=5pt,
    s sep=6pt,
    l sep=3mm,
    edge={draw=black!80},
  },
  for tree={
    if n children=0{
      draw=orange!25,
      rounded corners=3pt,
      fill=leafbg,
      node options={
        execute at begin node={\begin{varwidth}{8.5cm}\raggedright},
        execute at end node={\end{varwidth}}
      }
    }{
      if level=0{
        draw=purple!50,
        thick,
        rounded corners=4pt,
        fill=purple!6,
        font=\sffamily\bfseries,
        node options={
          execute at begin node={\begin{varwidth}{4.5cm}\centering},
          execute at end node={\end{varwidth}}
        }
      }{
        if level=1{
          draw=blue!40,
          thick,
          rounded corners=4pt,
          fill=blue!6,
          node options={
            execute at begin node={\begin{varwidth}{3.5cm}\raggedright},
            execute at end node={\end{varwidth}}
          }
        }{
          if level=2{
            draw=teal!40,
            thick,
            rounded corners=4pt,
            fill=teal!4,
            node options={
              execute at begin node={\begin{varwidth}{4.0cm}\raggedright},
              execute at end node={\end{varwidth}}
            }
          }{
            draw=violet!30,
            thick,
            rounded corners=4pt,
            fill=violet!4,
            node options={
              execute at begin node={\begin{varwidth}{4.5cm}\raggedright},
              execute at end node={\end{varwidth}}
            }
          }
        }
      }
    }
  }
  [Mitigation Strategies, s sep=30pt
    [Data-Level Mitigation\\
      \rev{\footnotesize Detailed in Figure~\ref{fig:data_mit_tree}},
      draw=blue!40,
      thick,
      rounded corners=4pt,
      fill=blue!6,
      node options={
        execute at begin node={\begin{varwidth}{3.5cm}\raggedright},
        execute at end node={\end{varwidth}}
      }
    ]
    [Model-Level Mitigation
      [\rev{Model Adaptation and Fine-Tuning}
        [{Package Hallucinations~{\scriptsize \cite{17}}, \rev{Large Language Models for Code}~{\scriptsize \cite{19}}, \rev{ReCode}~{\scriptsize \cite{74}}, \rev{CRPE}~{\scriptsize \cite{77}}, UCD-Training~{\scriptsize \cite{124}}, ShortCoder~{\scriptsize \cite{127}}}]
      ]
      [\rev{Reward-Guided Training}
        [{Codequal Analyzer~{\scriptsize \cite{52}}, REAL~{\scriptsize \cite{59}}, ReCode~{\scriptsize \cite{74}}}]
      ]
      [\rev{Training-Time Repetition Control}
        [{Repetition In Repetition Out~{\scriptsize \cite{85}}}]
      ]
    ]
    [Generation-Level Mitigation
      [Pre-Generation
        [\rev{Prompt Constraints}
          [{LLMs Meet Library Evolution~{\scriptsize \cite{2}}, \rev{SStuBs}~{\scriptsize \cite{16}}, \rev{Syntactic Robustness}~{\scriptsize \cite{32}}, Beyond Functional Correctness~{\scriptsize \cite{130}}}]
        ]
        [\rev{Privacy-Preserving Input Transformation}
          [{CodeCipher~{\scriptsize \cite{96}}}]
        ]
        [Retrieval-Augmented Generation
          [{Package Hallucinations~{\scriptsize \cite{17}}, CloudAPIBench~{\scriptsize \cite{27}}, AllianceCoder~{\scriptsize \cite{133}}, \rev{RustEvo\textsuperscript{2}}~{\scriptsize \cite{134}}, \rev{LLM Hallucinations in Practical Code Generation}~{\scriptsize \cite{137}}}]
        ]
      ]
      [In-Generation
        [Decoding-Time Intervention
          [{\rev{CodeIP}~{\scriptsize \cite{42}}, RPG~{\scriptsize \cite{84}}, Adadec~{\scriptsize \cite{131}}, RobGen~{\scriptsize \cite{135}}}]
        ]
      ]
      [Post-Generation
        [\rev{Feedback-Guided Regeneration}
          [{\rev{Package Hallucinations}~{\scriptsize \cite{17}}, \rev{Refining ChatGPT-Generated Code}~{\scriptsize \cite{69}}, MultiCodeIF~{\scriptsize \cite{129}}}]
        ]
        [\rev{Deterministic Repair and Transformation}
          [{\rev{LLMs Meet Library Evolution}~{\scriptsize \cite{2}}, Code Copycat Conundrum~{\scriptsize \cite{132}}}]
        ]
      ]
    ]
  ]
\end{forest}%
}
\caption{\rev{Multi-label, lifecycle-oriented taxonomy of strategies used to mitigate \codeissues. Read from left to right, the hierarchy locates interventions at the data, model, or generation level. The compact data-level node points to Figure~\ref{fig:data_mit_tree}; model-level strategies alter learned behavior, while generation-level strategies act before, during, or after a candidate is produced. A paper may appear in multiple leaves when it reports distinct interventions.}}
\Description{\rev{Taxonomy of generated-code issue mitigation strategies at the data, model, and generation levels, with corresponding literature references.}}
\label{fig:code_mit_tree}
\end{figure}

%% file: figures/data_mit_tree.tex
\begin{figure}[htbp]
\centering
\definecolor{leafbg}{HTML}{FFFDF0}
\resizebox{\textwidth}{!}{%
\begin{forest}
for tree={
    grow'=0,
    forked edges,
    anchor=west,
    child anchor=west,
    parent anchor=east,
    font=\sffamily\small,
    inner sep=5pt,
    s sep=6pt,
    l sep=3mm,
    edge={draw=black!80},
  },
  for tree={
    if n children=0{
      draw=orange!25,
      rounded corners=3pt,
      fill=leafbg,
      node options={
        execute at begin node={\begin{varwidth}{8.5cm}\raggedright},
        execute at end node={\end{varwidth}}
      }
    }{
      if level=0{
        draw=purple!50,
        thick,
        rounded corners=4pt,
        fill=purple!6,
        font=\sffamily\bfseries,
        node options={
          execute at begin node={\begin{varwidth}{4.5cm}\centering},
          execute at end node={\end{varwidth}}
        }
      }{
        if level=1{
          draw=blue!40,
          thick,
          rounded corners=4pt,
          fill=blue!6,
          node options={
            execute at begin node={\begin{varwidth}{3.5cm}\raggedright},
            execute at end node={\end{varwidth}}
          }
        }{
          if level=2{
            draw=teal!40,
            thick,
            rounded corners=4pt,
            fill=teal!4,
            node options={
              execute at begin node={\begin{varwidth}{4.0cm}\raggedright},
              execute at end node={\end{varwidth}}
            }
          }{
          }
        }
      }
    }
  }
  [Training Data Quality Issue Mitigation Strategies
    [Data Cleaning and Filtering
      [Execution-Feedback-Based Filtering
        [{Qwen2~{\scriptsize \cite{5}}, \rev{Qwen2.5}~{\scriptsize \cite{71}}, \rev{TeleChat2}~{\scriptsize \cite{72}}, Kimi K2~{\scriptsize \cite{73}}, Seed-Coder~{\scriptsize \cite{75}}, \rev{Code Llama}~{\scriptsize \cite{101}}}]
      ]
      [\rev{Deterministic Validation, Filtering, Sanitization, and Deduplication}
        [{Less Is More~{\scriptsize \cite{3}}, Qwen~{\scriptsize \cite{4}}, \rev{Qwen2}~{\scriptsize \cite{5}}, Quality In, Quality Out~{\scriptsize \cite{63}}, SwallowCode~{\scriptsize \cite{66}}, Qwen2.5~{\scriptsize \cite{71}}, \rev{TeleChat2}~{\scriptsize \cite{72}}, Seed-Coder~{\scriptsize \cite{75}}, DeepSeek-Coder~{\scriptsize \cite{79}}, StarCoder 2 and The Stack v2~{\scriptsize \cite{82}}, DataComp-LM~{\scriptsize \cite{98}}, RedStone~{\scriptsize \cite{100}}, Code Llama~{\scriptsize \cite{101}}, MG-Verilog~{\scriptsize \cite{112}}, \rev{aiXcoder-7B}~{\scriptsize \cite{117}}}]
      ]
      [\rev{LLM Scoring or Judging}
        [{Qwen~{\scriptsize \cite{4}}, Qwen2~{\scriptsize \cite{5}}, \rev{DataMan}~{\scriptsize \cite{6}}, Qwen3~{\scriptsize \cite{70}}, Qwen2.5~{\scriptsize \cite{71}}, \rev{TeleChat2}~{\scriptsize \cite{72}}, Kimi K2~{\scriptsize \cite{73}}, Seed-Coder~{\scriptsize \cite{75}}}]
      ]
      [\rev{Learned Quality or Relevance Scorers}
        [{\rev{Less Is More}~{\scriptsize \cite{3}}, \rev{DataMan}~{\scriptsize \cite{6}}, \rev{CIDRe}~{\scriptsize \cite{61}}, Seed-Coder~{\scriptsize \cite{75}}, DataComp-LM~{\scriptsize \cite{98}}}]
      ]
      [Manual Review and Curation
        [{Qwen~{\scriptsize \cite{4}}, \rev{TeleChat2}~{\scriptsize \cite{72}}, \rev{ReCode}~{\scriptsize \cite{74}}, StarCoder 2 and The Stack v2~{\scriptsize \cite{82}}}]
      ]
    ]
    [\rev{Mixture Design and Quality-Aware Sampling}
      [{Qwen~{\scriptsize \cite{4}}, DataMan~{\scriptsize \cite{6}}, Qwen3~{\scriptsize \cite{70}}, \rev{TeleChat2}~{\scriptsize \cite{72}}, Code Pretraining~{\scriptsize \cite{81}}, Code Llama~{\scriptsize \cite{101}}}]
    ]
    [Data Enhancement
      [\rev{LLM-Based Rewriting and Semantic Correction}
        [{\rev{SwallowCode}~{\scriptsize \cite{66}}, \rev{Qwen3}~{\scriptsize \cite{70}}, DataRecipe~{\scriptsize \cite{115}}}]
      ]
      [\rev{Rule- or Tool-Based Transformation}
        [{DataRecipe~{\scriptsize \cite{115}}}]
      ]
    ]
    [Data Augmentation
      [\rev{Curated-Source Expansion}
        [{\rev{StarCoder 2 and The Stack v2}~{\scriptsize \cite{82}}, Code Llama~{\scriptsize \cite{101}}}]
      ]
      [Synthetic Data Generation
        [{Qwen2~{\scriptsize \cite{5}}, Phi-4~{\scriptsize \cite{7}}, Qwen3~{\scriptsize \cite{70}}, Qwen2.5~{\scriptsize \cite{71}}, Kimi K2~{\scriptsize \cite{73}}, ReCode~{\scriptsize \cite{74}}, \rev{Seed-Coder}~{\scriptsize \cite{75}}, CRPE~{\scriptsize \cite{77}}, Code Llama~{\scriptsize \cite{101}}, MG-Verilog~{\scriptsize \cite{112}}, UCD-Training~{\scriptsize \cite{124}}, \rev{ShortCoder}~{\scriptsize \cite{127}}, APIKG4SYN~{\scriptsize \cite{128}}}]
      ]
    ]
  ]
\end{forest}%
}
\caption{\rev{Multi-label taxonomy of strategies used to mitigate \dataissues before or during dataset construction. Read from left to right, the hierarchy separates cleaning and filtering, mixture design, enhancement, and augmentation; it then distinguishes operational techniques and lists papers reporting them. A paper may appear in multiple leaves when it reports distinct dataset operations.}}
\Description{\rev{Taxonomy of training-data issue mitigation strategies, including deterministic and learned filtering, human curation, mixture design, rewriting, transformation, and dataset expansion, with corresponding literature references.}}
\label{fig:data_mit_tree}
\end{figure}

%% file: tables/mitigation_comparison.tex
\begin{table*}[!tp]
\centering
\caption{\rev{Comparison of mitigation families reported in the included papers. Outcomes remain tied to their original models, languages, datasets, analyzers, and metrics; the table supports conditional method selection rather than an overall ranking.}}
\label{tab:mitigation_comparison}
\begin{revisionblock}
\small
\renewcommand{\arraystretch}{1.12}
\setlength{\tabcolsep}{3pt}
\begin{tabular}{@{}>{\raggedright\arraybackslash}p{2.3cm} >{\raggedright\arraybackslash}p{3.7cm} >{\raggedright\arraybackslash}p{3.85cm} >{\raggedright\arraybackslash}p{4.6cm}@{}}
\toprule
\textbf{Family and lifecycle point} & \textbf{Applicable when} & \textbf{Operational advantage or reported outcome} & \textbf{Cost, trade-off, and evidence limitation} \\
\midrule
\textbf{Deterministic validation and filtering} (data construction) &
An undesirable property can be checked before training with a parser, rule, test, metadata field, or comparison set, and another training run is feasible. &
Removes invalid, vulnerable, duplicated, contaminated, or non-compliant material and can reduce later training work~\cite{3,63,82,117}. CleanTest reduced average fine-tuning time by 25.45\% and 10.00\% in two Java test datasets~\cite{3}. &
Rules and thresholds can remove useful samples and cover only encoded patterns. CleanTest had deteriorations in individual model--metric cells and Java-specific rules~\cite{3}; The Stack v2 reports a multi-stage curation pipeline but does not isolate each filter's downstream effect~\cite{82}. \\

\textbf{Scoring, sampling, and expansion} (data construction) &
Quality or domain labels, a target mixture, or a verified source of additional tasks is available. &
Supports graded selection rather than binary removal and can add underrepresented domains, languages, tasks, or solutions~\cite{6,77,82,101}. &
Labels, mixture targets, and source approval require human decisions; model scoring and synthetic generation add compute and can reproduce bias or duplication. DataMan's controlled results concern general text, and complete model pipelines do not isolate every mixture choice~\cite{6,70,101}. \\

\textbf{Rewriting and transformation} (data construction) &
Samples contain useful information but need style, interface, intent, self-containment, or algorithm changes rather than removal. &
Can retain source material while correcting an identified property. SwallowCode reports larger HumanEval-family gains from rewriting than from its scoring filter in a Python continual-pretraining setting~\cite{66}. &
Requires generation or refactoring plus validation. SwallowCode used 22\% more computation than scoring and its renaming caused an approximately 10-point MBPP decline; DataRecipe's aggregate tests found no significant overall improvement~\cite{66,115}. \\

\textbf{Model adaptation and reward-guided training} (training) &
Weights or prefixes can be updated and curated examples, tests, analyzers, or task-specific rewards match the deployment target. &
Amortizes a target behavior across later requests; reviewed examples address package use, security, maintainability, API migration, repetition, and output length~\cite{17,19,59,74,85,127}. &
Requires training compute and regression evaluation. Target improvements can accompany losses elsewhere: package-hallucination fine-tuning reduced HumanEval accuracy in its setting, and REAL's quality-only reward produced trivial non-functional outputs~\cite{17,59}. \\

\textbf{Prompt, retrieval, and decoding control} (generation) &
The model cannot be retrained or a constraint, current dependency, or white-box token signal is request-specific. &
Acts on the current request. Retrieval can supply evolving APIs, while grammar or checker-guided decoding can target repetition, robustness, a correctness--decoding-latency trade-off, or provenance~\cite{27,42,84,131,134,135}. &
Adds context, retrieval, or per-token cost. CloudAPIBench reports a 39.02-point high-frequency API decline with a suboptimal retriever~\cite{27}; RobGen requires logits and a checker and reports 33.4\%--197.9\% time overhead across its configurations~\cite{135}. \\

\textbf{Feedback-guided or deterministic repair} (post-generation) &
A compiler, test, analyzer, constraint checker, API map, or repetition rule returns a localized defect after generation. &
Reuses development tools to repair a candidate. Detailed diagnostics supported partial iterative repair~\cite{69}; MultiCodeIF reports higher constraint satisfaction over four rounds~\cite{129}; fixed mappings and patterns permit cheaper deterministic repair~\cite{2,132}. &
Repeated model calls and executions add cost, fixes can introduce another issue, and gains can plateau~\cite{69,129}. Deterministic repair covers only maintained mappings or patterns; function-level and synthetic benchmarks do not establish repository-level correctness. \\
\bottomrule
\end{tabular}
\end{revisionblock}
\Description{\rev{Comparison of deterministic data filtering, scoring and expansion, transformation, model training, generation-time control, and feedback-guided or deterministic repair by conditions, reported benefits, costs, trade-offs, and limitations.}}
\end{table*}

%% file: sections/discussion.tex
\section{Discussion}
\label{sec:discussion}

This section discusses the broader implications of the findings beyond the descriptive synthesis in Section~\ref{sec:findings}. We first distill several cross-cutting insights that connect \codeissues with \dataissues. We then summarize the main open challenges that continue to limit reliable quality governance in current research. Based on these observations, we outline a roadmap for data-centric quality governance and further position this perspective within the broader LLM engineering ecosystem.

\subsection{Cross-Cutting Insights on Code and Data Quality}

\begin{revisionblock}
\textbf{Training Data in Multi-Layer Quality Assurance.} RQ3 identifies recurring relationships between corpus properties and generated correctness, security, compliance, robustness, maintainability, understandability, efficiency, and generation derailment. Training data can be audited, filtered, balanced, and versioned before model training. Architecture, optimization and alignment, prompting, retrieval and tool use, decoding, and deployment context also shape generated code. A generated defect therefore motivates diagnosis across data, model, inference, and software-integration layers.
\end{revisionblock}

\rev{\textbf{Beyond Functional Correctness.} Functional correctness remains the dominant focus in the reviewed literature, but execution-based tests alone do not capture the non-functional and risk-related quality dimensions synthesized in RQ1. The reviewed studies use complementary security, structural, and human-centered assessments to evaluate these dimensions~\cite{20,22,83,93,130}. Evaluations should therefore combine functional tests with measures selected for the quality requirements of the target software.}

\begin{revisionblock}
\textbf{Context-Dependent Data Mixtures.} Real-world code corpora are skewed toward mainstream languages and common algorithmic tasks. The reviewed relationships involving prevalence imbalance and underexposure associate these distributions with uneven quality across languages, domains, task formats, and difficulty levels. Controlled mixture studies also show that rankings vary across benchmarks and settings. Data composition should therefore be selected and evaluated against the intended deployment domains rather than treated as a single universal balance.
\end{revisionblock}

\subsection{Open Challenges in Current Research}

\rev{\textbf{Benchmark Contamination and Evaluation Validity.} Training--evaluation overlap can inflate benchmark performance and obscure generalization to unfamiliar tasks~\cite{82,98}. The reviewed evidence includes controlled leakage experiments, similarity-based analyses, temporal comparisons, and reported potential overlap, with effects that vary across settings~\cite{80,93,109,120}. Contamination analysis should therefore report the type of membership evidence, the overlap criterion, and performance on matched uncontaminated or post-cutoff tasks.}

\rev{\textbf{Data Opacity and Temporal Drift.} Incomplete disclosure of training corpora limits independent analysis of provenance, benchmark overlap, and temporal coverage~\cite{80}. Software ecosystems also evolve as APIs are deprecated and vulnerabilities are repaired, creating version-sensitive evaluation conditions~\cite{2,107}. Versioned corpus records and time-aware evaluations can support diagnosis and remediation of outdated or insecure output patterns.}

\begin{revisionblock}
\textbf{Attribution across System Layers.} The RQ3 evidence ranges from controlled data interventions to corpus proxies, cross-model comparisons, and authors' analyses. Closed training corpora and bundled changes to data, model, and optimization often prevent separation of their individual contributions. Stronger attribution requires provenance records, controlled data ablations, influence estimation, and evaluations that hold model and inference conditions stable while varying a defined data property.
\end{revisionblock}

\subsection{\rev{Operational Guidance for Software Workflows}}

\begin{revisionblock}
RQ4 and RQ5 suggest a layered control sequence rather than a single universal quality gate. Deterministic checks can screen known syntax, dependency, security, provenance, duplication, and policy patterns first. Runnable changes can then receive project tests, relevant stress inputs, static analysis, and resource measurements. Model-based triage or feedback-guided regeneration is most useful for residual cases with a defined rubric or diagnostic, followed by contextual review for unresolved or high-consequence decisions. Every repaired candidate must re-enter the complete applicable gate because an intervention can introduce a different defect.

Table~\ref{tab:practitioner_workflow} maps these controls to dataset intake, dependency handling, local review, CI/CD, release, and repository maintenance. This mapping is our synthesis of separately evaluated controls; the included papers do not evaluate the complete workflow as an integrated production system. Projects must set thresholds and escalation rules for their languages, repositories, risk levels, and build environments, and retain artifact, model, prompt, retrieval, and tool versions for later diagnosis.
\end{revisionblock}

\input{tables/practitioner_workflow}

\subsection{A Roadmap for Data-Centric Quality Governance}

\rev{Future work can integrate data curation, model training, and post-generation repair across the model lifecycle. We identify three research directions:}

\begin{revisionblock}
\textbf{Data Provenance and Targeted Unlearning.} Training pipelines should preserve source, version, license, filtering, and transformation metadata to trace observed output patterns to relevant training data. Influence estimation and code-oriented unlearning can then test whether particular samples or corpus regions contribute to an observed output pattern and guide targeted sanitization without full retraining.
\end{revisionblock}

\rev{\textbf{Dynamic, Contamination-Resistant Benchmarks.} Static benchmarks can be complemented by regularly updated evaluations drawn from recent repositories. Combining execution tests with static analysis can assess functional correctness, complexity, and security while reducing prior-exposure risk.}

\rev{\textbf{Aligning Models with Software Engineering Standards.} Alignment objectives can incorporate signals from static analyzers, efficiency profilers, dependency-resolution tools, and test execution in addition to human preferences. Their effectiveness should be evaluated across functional and non-functional qualities.}

\subsection{Positioning Data-Centric Governance in the Evolving \llm Engineering Ecosystem}

\rev{Recent deployments increasingly embed models in retrieval pipelines, tool-augmented workflows, and multi-agent systems, often described as \textit{harness engineering}. These components provide additional points for managing code quality.}

\begin{revisionblock}
Harness-level engineering manages quality at deployment time. Retrieval can supply current or domain-specific knowledge, while compilers, tests, static analyzers, and policy checks can intercept insecure or non-executable outputs. Data-centric controls address a different stage by improving the patterns and coverage presented during training. Combining these layers provides both preventive curation and runtime verification.
\end{revisionblock}

\begin{revisionblock}
Some capabilities previously handled by external agents, including tool use, iterative debugging, and multi-step planning, are increasingly learned during pre-training or post-training. Their examples and feedback data become additional quality-control targets, while external orchestration continues to shape execution. Reliable systems therefore require coordinated governance of training data, model behavior, and the surrounding engineering harness.
\end{revisionblock}

%% file: tables/practitioner_workflow.tex
\begin{table*}[!t]
\centering
\caption{\rev{Author-synthesized integration of reviewed controls into software workflows. The cited papers evaluate individual controls, not the complete CI/CD arrangements shown here. Uncited workflow extensions are our recommendations and require project-specific validation and thresholds.}}
\label{tab:practitioner_workflow}
\begin{revisionblock}
\small
\renewcommand{\arraystretch}{1.15}
\setlength{\tabcolsep}{3pt}
\begin{tabular}{@{}>{\raggedright\arraybackslash}p{2.3cm} >{\raggedright\arraybackslash}p{6.6cm} >{\raggedright\arraybackslash}p{5.75cm}@{}}
\toprule
\textbf{Workflow checkpoint} & \textbf{Recommended control} & \textbf{Decision and scope condition} \\
\midrule
\textbf{Dataset intake and model update} &
Record source and snapshot or version identifiers; apply license and opt-out policies; detect exact and near duplicates; redact PII; scan for malware; and use version history to flag potentially outdated, buggy, or vulnerable files~\cite{82,107,119}. &
Retain a review path for uncertain provenance, licenses, and destructive filters. Preserve dataset and filter versions so model evaluations can be related to the actual training snapshot. \\

\textbf{Generation context and dependencies} &
Declare and resolve the target dependency version. For deprecated, low-frequency, or post-cutoff APIs, supply version-specific documentation or examples~\cite{2,27,134}. &
Use selective retrieval and validate retrieved material: CloudAPIBench reports a 39.02-percentage-point drop for high-frequency APIs under one suboptimal retrieval setting~\cite{27}. Confirm compatibility in the target build. \\

\textbf{Local and code-review gate} &
Build generated changes and run them in an isolated test environment. Use project tests and relevant stress inputs, supported-language linters and insecure-practice rules, and resource measurements when efficiency matters~\cite{20,32,69,138}. &
Route failures and ambiguous flags to the author or reviewer. Finite differential tests do not prove equivalence~\cite{32}, and CyberSecEval's rules cover specified insecure practices rather than every vulnerability~\cite{20}; architecture and organizational requirements remain contextual decisions. \\

\textbf{CI/CD and release gate} &
Run generated changes through the same build and release gates as human changes. The included papers directly support compiler, test, and static-analysis feedback~\cite{69}, insecure-practice checks~\cite{20}, and version-specific API checks~\cite{2,134}; repository-specific dependency and license gates are our workflow synthesis. &
In one four-round Java/Python LeetCode study, gains concentrated in rounds 1--2 and repairs sometimes introduced new issues~\cite{69}. We therefore recommend a retry limit, a complete gate after each attempt, and retention of model, prompt, context, and tool versions in the build record. \\

\textbf{Repository maintenance and feedback} &
Track supported dependency and API versions and refresh version-specific retrieval sources and evaluation tasks as ecosystems evolve~\cite{2,134}. Triage confirmed test, security, and review findings as candidates for rules, evaluation sets, or curated feedback data; promote data only after the intake controls above. &
Do not treat production correlation as training-data attribution. Changes to a model or corpus require controlled reevaluation against the intended languages, repositories, and quality attributes. \\
\bottomrule
\end{tabular}
\end{revisionblock}
\Description{\rev{Author-synthesized mapping from reviewed controls to dataset intake, dependency handling, local review, continuous integration and delivery, and repository maintenance, with scope conditions and the boundary between evaluated controls and workflow recommendations.}}
\end{table*}

%% file: sections/threats.tex
\section{Threats to Validity}
\label{sec:threats}

\begin{revisionblock}
\textbf{Internal Validity.} Despite a multi-database search and backward and forward snowballing, relevant papers may be missed because of indexing delays, evolving preprints, or terminology differences between software engineering and artificial intelligence. RQ3 also requires interpretive judgments when a paper combines several data changes or reports several outcomes. We reduced this subjectivity by applying the RQ1 and RQ2 definitions consistently, requiring each training-data condition and generated outcome to be supported within the same paper, and consolidating overlapping descriptions. The 12 families nevertheless depend on qualitative categorization.
\end{revisionblock}

\textbf{Construct Validity.} The concepts of code and dataset quality are multifaceted and lack universally standardized definitions. While our taxonomy synthesizes existing software quality frameworks and recent literature, this abstraction may oversimplify nuanced dimensions like semantic alignment or safety constraints. Furthermore, evaluating included papers using our 16-point quantitative matrix (Section~\ref{sec:methodology}) relies on the transparency of the original publications; sparse methodological reporting in certain papers may introduce scoring inconsistencies. As the field rapidly evolves, emerging quality attributes may transcend the constructs formalized in this review.

\begin{revisionblock}
\textbf{External Validity.} The \numstudies included papers predominantly examine English-language corpora, mainstream programming languages, open-source data, and current \llm architectures and training regimes. The mapping synthesis may therefore differ for low-resource languages, proprietary enterprise codebases, future model families, or deployment pipelines with different retrieval and validation controls.
\end{revisionblock}

\begin{revisionblock}
\textbf{Conclusion Validity.} Heterogeneous tasks, models, interventions, and outcome measures preclude a formal meta-analysis. The family frequencies summarize literature coverage, and the reported effect sizes are not directly comparable. Mixed and small findings temper several associations, while publication bias may favor successful interventions and novel frameworks over neutral results. We therefore draw association-level conclusions unless a primary empirical study isolates a causal effect.
\end{revisionblock}

%% file: sections/conclusion.tex
\section{Conclusion}
\label{sec:conclusion}

\begin{revisionblock}
This systematic review of \numstudies included papers establishes taxonomies of generated-code and training-data quality issues and identifies 12 recurring relationship families between them. Research is concentrated on correctness, underexposure, prevalence imbalance, defective training code, and supervision quality, whereas structural quality, clarity, and robustness remain less studied. The evidence ranges from controlled data interventions to corpus audits, cross-model comparisons, reviews, and authors' analyses.

The findings position training data as one actionable layer in code-generation quality assurance. Reliable development tools also depend on model training and alignment, inference-time context and decoding, external validation, and integration with software-engineering workflows. End-to-end quality pipelines can combine provenance and targeted curation with dynamic evaluation, static analysis, testing, dependency checks, and post-generation repair. This coordinated approach connects data management with the functional and non-functional standards expected of production software.
\end{revisionblock}